\documentclass[a4paper,amsmath,amssymb,twocolumn,superscriptaddress]{revtex4}
\usepackage{doi}
\usepackage{hyperref}
\hypersetup{colorlinks=true,linkcolor=blue,
  citecolor=cyan,}
\usepackage{dcolumn}
\usepackage{bm}
\usepackage{xcolor}
\usepackage{amsmath}
\usepackage{amssymb}
\usepackage[T1]{fontenc}
\usepackage{lmodern}
\usepackage{subfigure}
\usepackage{textcomp}
\usepackage{newunicodechar}

\newcommand{\beq}{\begin{equation}}
\newcommand{\eeq}{\end{equation}}
\newcommand{\bea}{\begin{eqnarray}}
\newcommand{\eea}{\end{eqnarray}}

\usepackage{float, geometry,lipsum}
\usepackage{graphicx}
\usepackage{dcolumn}
\usepackage{bm}
\usepackage{color}
\usepackage{enumitem}
\usepackage{amsmath}
\usepackage{amssymb}
\newunicodechar{−}{\ensuremath{-}}

\begin{document}

\title{\bf QPO-like Signatures and Hydrodynamical Variability in Accretion around a JNW-type Compact Spacetime in Freund-Nambu Scalar-Tensor Gravity}

\author{Orhan~Donmez}
\email{orhan.donmez@aum.edu.kw}
\affiliation{College of Engineering and Technology, American University of the Middle East, Egaila 54200, Kuwait}

\author{M. Yousaf}
\email{myousaf.math@gmail.com}
\affiliation{Department of Mathematics, Virtual University of Pakistan, 54-Lawrence Road, Lahore 54000, Pakistan.}

\author{G. Mustafa}
\email{gmustafa3828@gmail.com}
\affiliation{Department of Physics, Zhejiang Normal University,
Jinhua 321004, China}

\begin{abstract}
scalar tensor theories of gravity provide a broad as well as physically rich extension of general theory of relativity (GTR) by allowing the gravitational interaction to be mediated not only by the spacetime metric but also by scalar degrees of freedom. In this manuscript, we present a new exact solution in the Freund-Nambu scalar-tensor (FNST) gravity scenario, representing a nontrivial scalar-tensor generalization of the Janis-Newman-Winicour (JNW) naked-singularity geometry, characterized by an additional coupling parameter $q$ in the scalar sector. 
We also numerically solve the general relativistic hydrodynamic (GRH) equations in order to investigate the shock-cone mechanism formed by Bondi-Hoyle-Lyttleton (BHL) accretion around this compact spacetime on the equatorial plane.  Depending on the value of the deformation parameter $n$, we show that stronger scalar-tensor deviations modify the shock-cone morphology, significantly increase the amount of matter accumulated near the central compact object, and enhance the oscillatory behavior of the shock cone. The Lorentzian-like peaks obtained from the numerically computed power spectral density (PSD) are interpreted as hydrodynamically generated quasi-periodic-oscillation (QPO)-like modes. These modes are driven by shock cone oscillations and by the compression and rarefaction of the plasma trapped inside the cone. Finally, for a compact object with mass parameter $M=10M_{\odot}$, the numerically extracted frequencies are found mainly in the range from a few Hz up to approximately $100$ Hz. These frequencies overlap with the QPO ranges reported in stellar-mass black-hole-candidate systems. In particular, the frequencies obtained for the FNST2--FNST4 models fall within the range of timing features reported for the source GRS 1915+105. These results suggest that the exterior hydrodynamical variability of FNST compact spacetimes may provide phenomenological diagnostics of scalar-field-induced deviations from the Schwarzschild reference case.
\end{abstract}

\maketitle

\date{\today}

\section{Introduction}\label{s1}

Einstein's GTR provides a successful geometric description of gravitation, in which the gravitational field is encoded in the spacetime metric. Since its formulation, GTR passed a wide range of observational as well as experimental tests and remains the standard scenario for describing gravitational phenomena from Solar System scales to compact astrophysical entities. Nevertheless, several fundamental issues, including the nature of dark energy, the origin of cosmic acceleration, the behavior of gravity in the strong field regime, as well as the possible resolution of spacetime singularities, continue to motivate the study of extended theories of gravity. Among the most widely investigated alternatives are scalar tensor theories as studied in~\cite{BransDicke1961,FujiiMaeda2003,SotiriouFaraoni2010,Clifton2012}. Such theories provide a natural theoretical setting for exploring deviations from GTR, possible variations of the effective gravitational coupling, as well as modifications of gravitational dynamics on both cosmological and astrophysical scales. The development of scalar tensor gravity can be viewed as a modern continuation of early attempts to describe gravitation through scalar fields, although contemporary scalar tensor theories differ substantially from the earlier scalar models. In these theories, the scalar field is not introduced as a replacement for the metric structure of spacetime; rather, it appears as an additional dynamical component coupled to the geometric sector, while this feature allows scalar tensor gravity to preserve the geometric foundation of GTR while introducing new physical effects through nonminimal coupling, scalar self interactions, or direct coupling with matter fields. Such extensions extensively used in cosmology, compact entity physics, as well as strong gravity phenomenology~\cite{FujiiMaeda2003,DeFelice2010,Doneva:2022ewd,Berti2015,Barack2019}. In particular, scalar tensor models may account for accelerated cosmic expansion without invoking a fundamental cosmological constant, since the scalar field can dynamically contribute to the effective dark energy sector~\cite{Copeland2006,AmendolaTsujikawa2010}.

An early as well as important realization of this class of theories is the FNST theory, originally proposed in the context of scalar field extensions of gravitation~\cite{Freund1968PR}, whereas in this scenario, gravity is mediated not only by the spacetime metric but also by a dynamical scalar field, which contributes an additional degree of freedom to the gravitational interaction. The presence of this scalar field can modify the structure of spacetime and influence the properties of compact entities, accretion flows, and particle dynamics in strong gravitational fields. The gravitational action in the Freund Nambu (FN) scenario generally leading to coupled nonlinear field equations obtained through variation with respect to the metric and scalar degrees of freedom. Due to these features, the FN theory provides a useful theoretical laboratory for examining departures from GTR and for testing scalar field effects in relativistic astrophysical environments~\cite{FujiiMaeda2003,DeFelice2010,Clifton2012}. Scalar field configurations are also closely connected with the appearance of non Schwarzschild compact geometries, while a particularly relevant example is provided by static and spherically symmetric solutions sourced by scalar fields, where the central singularity may not be hidden behind an event horizon. In GTR, such a configuration is commonly referred to as a naked singularity, whereas mathematically, spacetime singularities are characterized by divergent curvature invariants, such as the Kretschmann scalar, which indicate the breakdown of the classical spacetime description. A naked singularity can in principle communicate with distant observers through causal geodesics, however one of the well known scalar field solutions of this type is the JNW spacetime~\cite{Janis68}, which generalizes the Schwarzschild geometry by incorporating a massless scalar field. The additional scalar parameter significantly changes the geodesic structure, effective potential, as well as orbital properties of test particles, thereby producing potentially observable differences from the standard black hole (BH) scenario.

The motion of test particles around compact entities is a powerful tool for probing the underlying spacetime geometry, while in static as well as spherically symmetric backgrounds, the analysis of circular orbits and some other features provide valuable information about the gravitational field in the strong field regime~\cite{stuchlik2013,bambi2017,Turakhonov:2024smp,Turakhonov:2024xfg,Umarov:2025wzm,Turimov:2024hwh,Boboqambarova:2021cbf,Turimov:2021jgk,Turimov:2020fme}. These quantities are directly related to observational signatures arising from accretion disks around compact entities, while in particular, QPOs as well as compact binaries, are widely regarded as sensitive probes of the orbital and epicyclic motion of matter in the vicinity of strong gravitational sources~\cite{2024ChJPh..92..143R,2025PDU....5002102R}. Since QPO frequencies depend strongly on the spacetime geometry, their study provides a promising way to distinguish between GTR BHs, scalar field compact entities, and modified gravity solutions. To explain the origin of high frequency QPOs there are some theoretical scenarios, whereas among them, the epicyclic resonance model received considerable attention because it relates the observed twin peak QPOs to combinations of the orbital as well as epicyclic frequencies of nearly circular motion in the accretion disk~\cite{torok2005,stuchlik2008}. In this approach, the upper as well as lower QPO frequencies or with specific resonant combinations of these quantities. Since the epicyclic frequencies are determined by the background metric, any modification induced by the FN scalar field may leave detectable imprints on the QPO spectrum. Therefore, the investigation of QPO signatures in the FN spacetime offers a physically meaningful route to constrain scalar tensor modifications of gravity through astrophysical observations.

In addition to geodesic motion, the behavior of plasma in compact entities is of central importance for understanding accretion phenomena, while realistic accretion disks are composed of ionized matter interacting with strong gravitational fields, as well as the presence of plasma may influence wave propagation, disk oscillations, instability growth, and the observed electromagnetic spectrum. Hydrodynamical variability can play an important role in angular momentum transport, energy dissipation, as well as the excitation of oscillation modes in accretion flows. In modified gravity backgrounds, such as the spherically symmetric vacuum solution of FNST gravity, the scalar degree of freedom may alter the effective potential and characteristic frequencies associated with plasma motion. Consequently, the combined analysis of QPOs and hydrodynamical variability can provide deeper insight into the physical viability of scalar tensor compact entity geometries. In this direction, horizon free compact configurations also investigated as possible alternatives to BHs, with particular emphasis on their stability properties and electromagnetic signatures in~\cite{yousaf2025implications}, while accretion processes around compact entities likewise attracted considerable attention, as nonlinear electrodynamics as well as alternative gravity models can significantly modify the surrounding BH environment~\cite{donmez2026relativistic,donmez2026accretion}. Some related studies of high frequency QPOs in BH binaries, and BH shadows as well as compact entities phenomenology in different gravitational backgrounds to strengthened the role of astrophysical observations in distinguishing between GTR and its extensions~\cite{belloni2012high,johannsen2013photon}. To confront theoretical predictions with observational data, reliable statistical techniques are required, while specific methods are widely used for parameter estimation because they allow an efficient exploration of multidimensional parameter spaces and provide posterior probability distributions for model parameters~\cite{sharma2017,foreman-mackey2013,Shabbir:2026qlh,Shermatov:2025rpj,Turimov:2024orr}. In BH astrophysics some specific sources used to estimate compact entity parameters as well as to test deviations from standard GTR based metrics studied in~\cite{shafee2006,kolos2023,Hoshimov:2025tdx}. Recent investigations have shown that combining geodesic analysis, QPO modeling, shock morphology, PSD signatures, as well as statistical parameter estimation can place meaningful constraints on modified gravity models and alternative compact entity solutions~\cite{stuchlik2021,bambi2018,donmez2026disformal} and probing of rotating Kerr-Bertotti-Robinson BHs and fuzzy BH in the background of modified gravities presented in \cite{yousaf2024fuzzy,singh2026probing}.

In the present study, the FNST scenario is considered in its vacuum as well as massless scalar field configuration in order to isolate the purely geometric effects of the scalar degree of freedom, while in this setting, the matter sector and scalar field mass contribution are neglected, allowing the analysis to focus on the influence of the scalar field on the background spacetime. The resulting static and spherically symmetric solution contains a deformation parameter associated with the scalar field contribution, whereas in the appropriate limiting case, this geometry recovers the well known scalar field solution of JNW type, while the Schwarzschild spacetime is obtained when the scalar field deformation is removed and related work is presented in~\cite{Freund1968PR,Janis68}. Therefore, this background provides a suitable arena for examining how scalar tensor corrections modify strong field dynamics around compact entites. To investigate the accretion properties of this spacetime, we analyze the hydrodynamical evolution of plasma through BHL-type accretion in the equatorial plane, while the numerical treatment is based on general relativistic hydrodynamical evolution on a fixed FNST background, where the accreting matter is modeled as a perfect fluid and evolved using high resolution shock capturing techniques. Such an approach is well suited for resolving discontinuities, shock fronts, and matter compression regions that naturally arise in supersonic accretion flows around compact entities~\cite{Donmez2004ASS,Donmez2006AMC,Donmez2012MNRAS,Donmez2024JCAP}. In this configuration, matter injected from the upstream region is gravitationally focused toward the compact entity, whereas the dynamical behavior of this cone provides an important hydrodynamical study of the underlying spacetime geometry. The shock cone morphology obtained in the FNST models exhibits a clear dependence on the scalar field deformation parameter, although the overall formation of a downstream shock cone remains similar to the Schwarzschild case, the internal density distribution, opening angle, near-inner-boundary matter accumulation are modified as the deformation from Schwarzschild geometry becomes stronger. In particular, stronger scalar tensor deviations enhance the density of plasma trapped inside the shock cone as well as produce a more extended high density wake behind the compact entity. 
This behavior indicates that the scalar-field-induced modification of the spacetime affects the gravitational focusing of the incoming material. Consequently, the shock-cone structure may serve as a hydrodynamical diagnostic for distinguishing FNST naked-singularity or horizonless compact-object geometries from the Schwarzschild black-hole reference case~\cite{Donmez2024MPLA}. 

The time evolution of the rest-mass flux through the inner excision boundary provides a direct link between the simulated exterior hydrodynamical flow and possible observational variability, whereas in the FNST models, this measured mass flux shows persistent oscillatory behavior even after the system approaches a quasi-stationary configuration. These oscillations arise from the time-dependent motion of the shock cone and from compression and rarefaction of the trapped plasma. As the scalar-tensor deformation increases, both the average mass flux through the inner excision boundary and its temporal variability become stronger. This suggests that the deformation parameter can leave observable imprints not only on the amount of matter transported through the inner excision boundary but also on the amplitude and variability pattern of the emitted radiation. The PSD analysis of the inner-boundary mass-flux signal further reveals that the shock-cone oscillations can generate QPO-like frequency components. The resulting spectra contain several Lorentzian like peaks whose positions, coherence, as well as spectral strengths vary from the Schwarzschild model to the different FNST configurations, while some of these modes fall within the frequency ranges commonly associated with low frequency and intermediate/high frequency QPOs observed in stellar mass BH systems. In particular, the model frequencies can be compared with timing features reported in some particular sources after taking into account the appropriate mass scaling~\cite{Rodriguez:2004kg,Varniere:2018zea,Homan:2004pp}. Consequently, the present work investigates QPO signatures as well as hydrodynamical oscillations in accretion around a spherically symmetric vacuum solution in FNST gravity, while we examine the influence of the scalar field parameter on the orbital motion of test particles, the location of stable circular orbits, and the associated epicyclic frequencies. This manuscript is arranged as in section~\ref{Sec:Formulations}, we briefly present the basics of FNST gravity as well as discuss the corresponding field equations, while section~\ref{Sec:ScalarF} is devoted to the derivation of the massless type solution and its connection with the static spherically symmetric background considered. In Section~\ref{Num1}, we investigate the hydrodynamical accretion dynamics of plasma in the FNST spacetime, and in particular, subsection~\ref{Num2} analyzes the shock cone morphology as well as azimuthal density distribution, while subsection~\ref{Num3} discusses the oscillations of the inner-boundary mass flux as well as the corresponding PSD signatures, however section~\ref{compare} compares the hydrodynamically generated QPO like modes with the observed timing ranges reported for accreting black-hole-candidate systems. Finally, the main results as well as concluding remarks are summarized in section~\ref{Concl}.

\section{FNST gravity\label{Sec:Formulations}}

In the framework of FNST gravity action can be written in geometrized units $(G=c=1)$~\cite{Freund1968PR} as:
\begin{align}\label{action}
{\cal S} = \frac{1}{16\pi} \int d^4x, \sqrt{-g} \left(R - \frac{2\partial_\eta\varphi\partial^\eta\varphi}{1 + 2q\varphi} + 2\xi^2 \varphi^2 \right) + {\cal S}_M \ ,
\end{align}
Here, $g=|g_{\eta\zeta}|$, however $R$ and $\varphi$ denote the Ricci scalar as well as a massive real scalar field with mass $\xi$, respectively, while the parameter $q$ measures the scalar geometry coupling, whereas ${\cal S}_M$ represents the matter action coupled to $(1+2q\varphi)g_{\eta\zeta}$ which is conformally rescaled metric, therefore, the mat(esector ta+rkes the form
${\cal S}_M=\left((2q\varphi+1)g_{\eta\zeta},Q_M\right){\cal S}_M.$
So, varying \eqref{action} w.r.t. $g_{\eta\zeta}$ and $\varphi$ yields
\begin{align}\label{eq}
&G_{\eta\zeta}=R_{\eta\zeta}-\frac{1}{2}g_{\eta\zeta}R=T_{\eta\zeta}+T^M_{\eta\zeta} \ , \\
\label{eqf}
&\varphi(\square - \xi^2) = \left( T^M + T \right)q \ ,
\end{align}
Here, $G_{\eta\zeta}$ represents the Einstein tensor, as well as $\square\equiv -\nabla^\eta\nabla_\eta$ denotes the covariant D'Alembertian operator, while the stress energy tensor corresponding to the scalar field sector can then be written in the following form as
\begin{align}
T_{\eta\zeta} = \frac{2\partial_\eta\varphi\partial_\zeta\varphi}{1 + 2q\varphi} - g_{\eta\zeta} \left( \frac{\partial_\gamma\varphi,\partial^\gamma\varphi}{1 + 2q\varphi} - \xi^2 \varphi^2 \right) \ ,
\end{align}
with its trace given by
\begin{align}
T = g^{\eta\zeta} T_{\eta\zeta} = -\frac{2\partial_\eta\varphi\partial^\eta\varphi}{1 + 2q\varphi} + 4\xi^2 \varphi^2 \ .
\end{align}
In this study, we derive an exact analytical solution of the Einstein field equations generated by a massless scalar field in the FN scenario. Accordingly, we focus on the vacuum configuration by imposing $\xi=0$ and ${\cal S}_M=0$, thereby eliminating the scalar field mass term as well as the ordinary matter contribution, while this choice enables a direct investigation of the scalar field's geometric role in shaping the curved spacetime. The associated field equations and the resulting exact solution are presented in the next section.

\section{Exact Solution for a Massless Scalar Field}\label{Sec:ScalarF}

In the absence of the ordinary matter sector, the dynamical equations of the Einstein massless scalar field system in the FN scenario take the following form:

\begin{align}
\square\varphi=-\frac{q}{1+2q\varphi}\,
\partial_\eta\varphi\,\partial^\eta\varphi .
\end{align}
Equivalently, the massless scalar field equation can be written in the covariant form as
\begin{align}\label{KG}
\frac{1}{\sqrt{-g}}\partial_\xi
\left(\sqrt{-g}\,\partial^\xi\varphi\right)
=
\frac{q}{1+2q\varphi}\,
\partial_\eta\varphi\,\partial^\eta\varphi .
\end{align}
Let us, we now assume $\varphi=\varphi(r)$. Under this assumption, Eq.~\eqref{KG} becomes
\begin{align}
\frac{1}{\sqrt{-g}}\frac{d}{dr}
\left(\sqrt{-g}\,g^{rr}\frac{d\varphi}{dr}\right)
=
\frac{q}{1+2q\varphi}\,
g^{rr}\left(\frac{d\varphi}{dr}\right)^2 .
\end{align}
Introducing the redefined scalar function $F=1+2q\varphi$, so we get
\begin{align}
\left(\sqrt{-g}g^{rr}\frac{dF}{dr}\right)^{-1}\frac{d}{dr}
\left(\sqrt{-g}g^{rr}\frac{dF}{dr}\right)=\frac{1}{2F}\left(\frac{dF}{dr}\right)\ ,
\end{align}
implies
\begin{align}
&\frac{d}{dr}\ln\left(\sqrt{-g}g^{rr}\frac{dF}{dr}\right)=\frac{d}{dr}\ln\sqrt{F}\ ,
\end{align}
and one obtains by performing some basic mathematical exercise
\begin{align}
&\frac{dF}{\sqrt{F}}=\frac{2C_1dr}{\sqrt{-g}g^{rr}}\ ,
\end{align}
or
\begin{align}\label{solF}
&\sqrt{F}=\int\frac{C_1dr}{\sqrt{-g}g^{rr}}+C_2\ ,
\end{align}
where $C_1$ and $C_2$ are constants of integration. Thus, the static and spherically symmetric spacetime solution in FN theory takes the form \cite{Davlataliev:2026vkx}
\begin{align}\label{JNW}
ds^2&=-\left(1-\frac{2M}{nr}\right)^{n}dt^2
+\left(1-\frac{2M}{nr}\right)^{-n}dr^2\nonumber\\
&+r^2\left(1-\frac{2M}{nr}\right)^{1-n}
\left(d\theta^2+\sin^2\theta d\phi^2\right),
\end{align}
where $M$ denotes the gravitational mass of the compact object, while the dimensionless parameter $n$ characterizes the contribution arising from the scalar field sector. The corresponding scalar field is obtained as
\begin{align}
\varphi
=\frac{\sqrt{1-n^2}}{2}
\ln\left(1-\frac{2M}{nr}\right)
\left[
1+\frac{q}{2}\frac{\sqrt{1-n^2}}{2}
\ln\left(1-\frac{2M}{nr}\right)
\right].
\end{align}
In the limit of the standard Einstein scalar field system within GTR, the FN coupling parameter vanishes, while under this restriction, the metric \eqref{JNW} reduces to JNW naked singularity spacetime~\cite{Janis68}, whereas scalar field becomes

\begin{align}
\varphi=\frac{\sqrt{1-n^2}}{2}\ln\left(1-\frac{2M}{nr}\right)\ .    
\end{align}

The causal and singularity structure of the metric must be interpreted carefully. The metric coefficient contains the factor $A(r)=1-\frac{2M}{nr}$ which vanishes at $r_s=\frac{2M}{n}$.  For $n=1$, the scalar field vanishes and the metric reduces to the Schwarzschild black-hole solution. In this special case, $r_s=2M$ is the regular event horizon. However, for $0<n<1$, the scalar field does not vanish and contains a logarithmic divergence at $r=r_s$. Moreover, the areal radius of the two-spheres is
$R(r)=r\left(1-\frac{2M}{nr}\right)^{(1-n)/2}$, which approaches zero as $r\rightarrow r_s$. Therefore, $r=r_s$ corresponds to a physical curvature singularity rather than a regular event horizon. The spacetime is consequently of JNW type and represents a naked-singularity or horizonless compact-object geometry. In the following hydrodynamical analysis, only the Schwarzschild case $n=1$ is interpreted as a black hole. The models with $0<n<1$ are interpreted as FNST compact-object spacetimes, and the numerical inner boundary is placed outside $r_s$ as an excision boundary.

\begin{figure*}
\centering
\includegraphics[width=7.0cm,height=5.0cm]{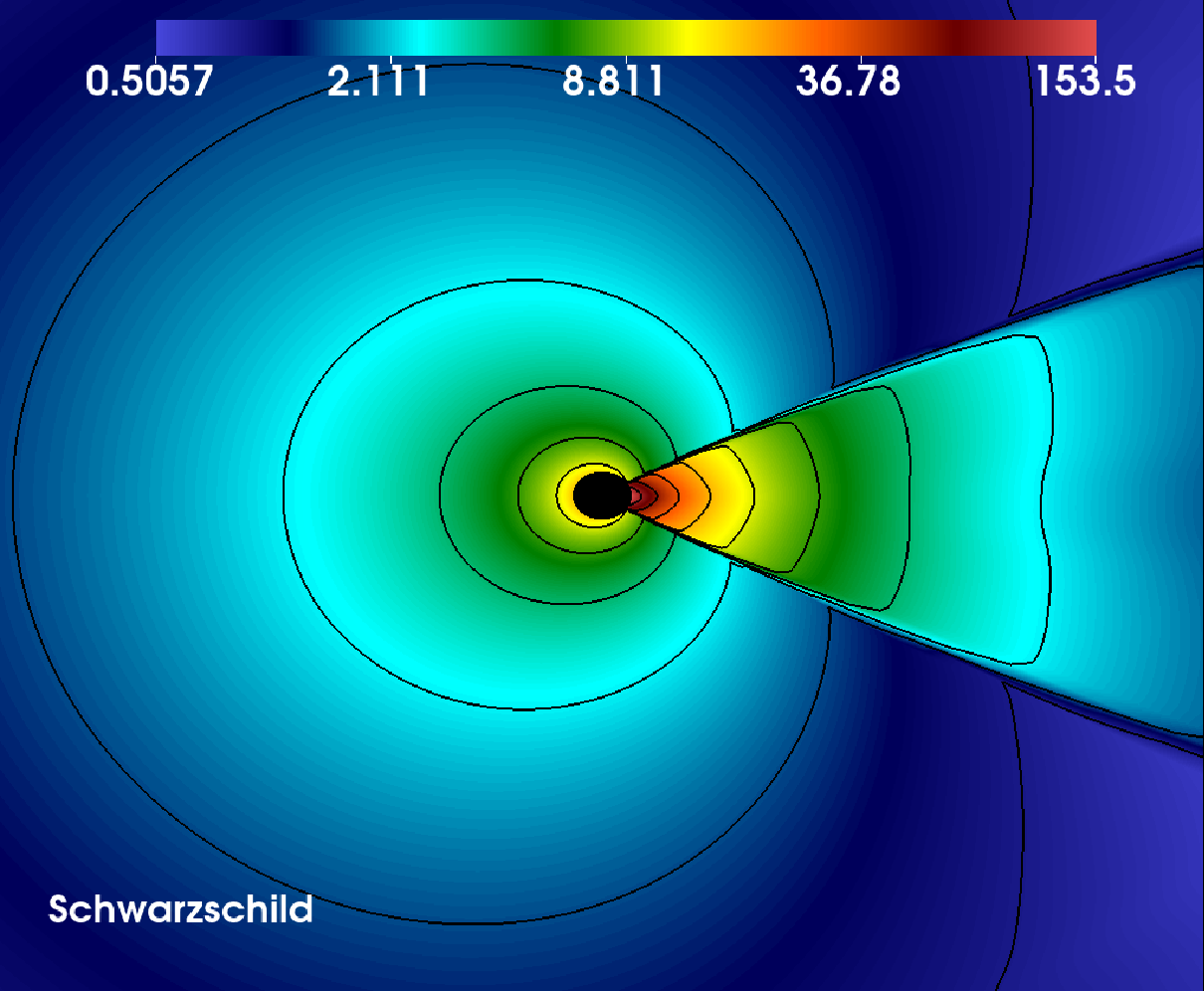}
\includegraphics[width=7.0cm,height=5.0cm]{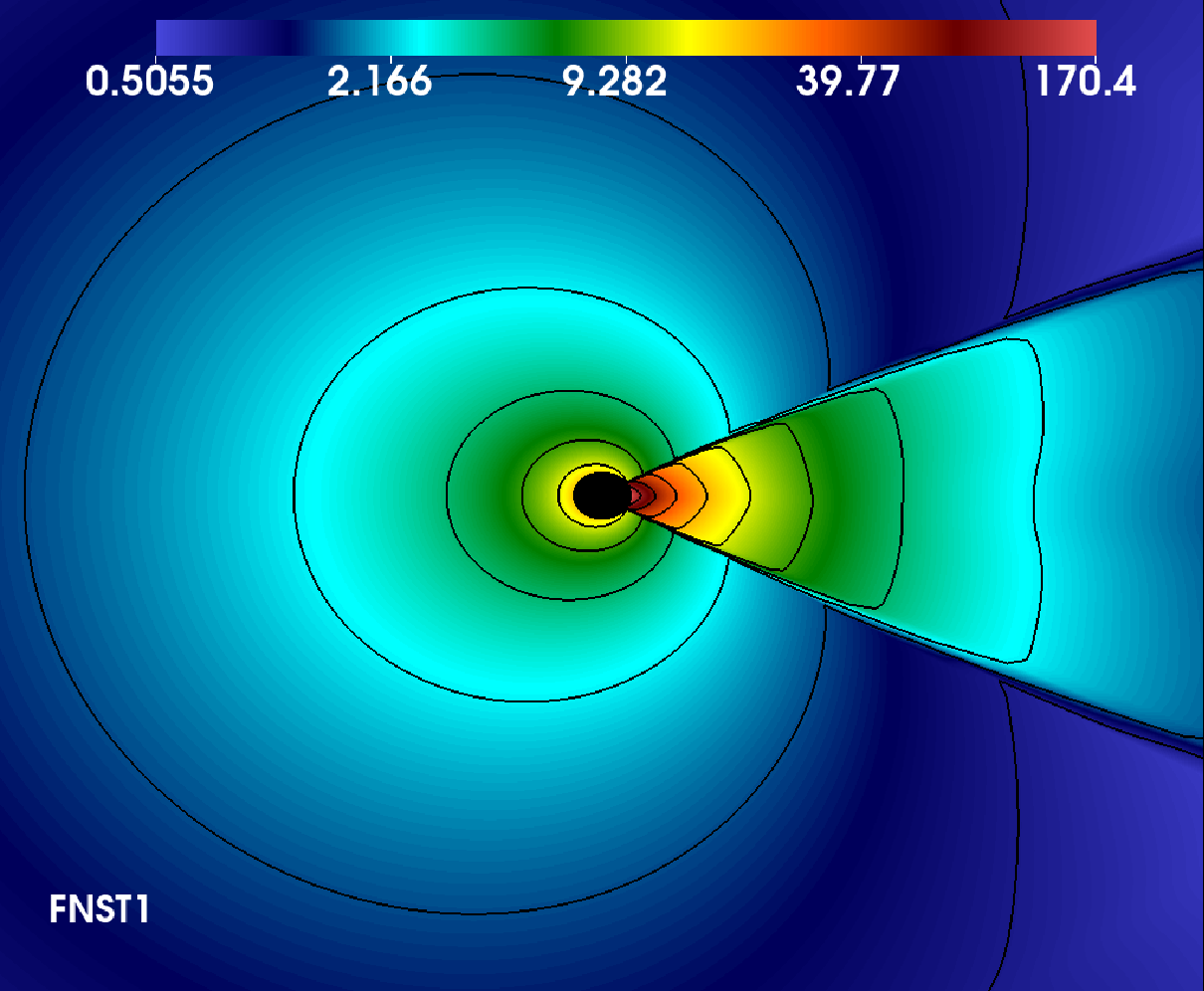}
\includegraphics[width=7.0cm,height=5.0cm]{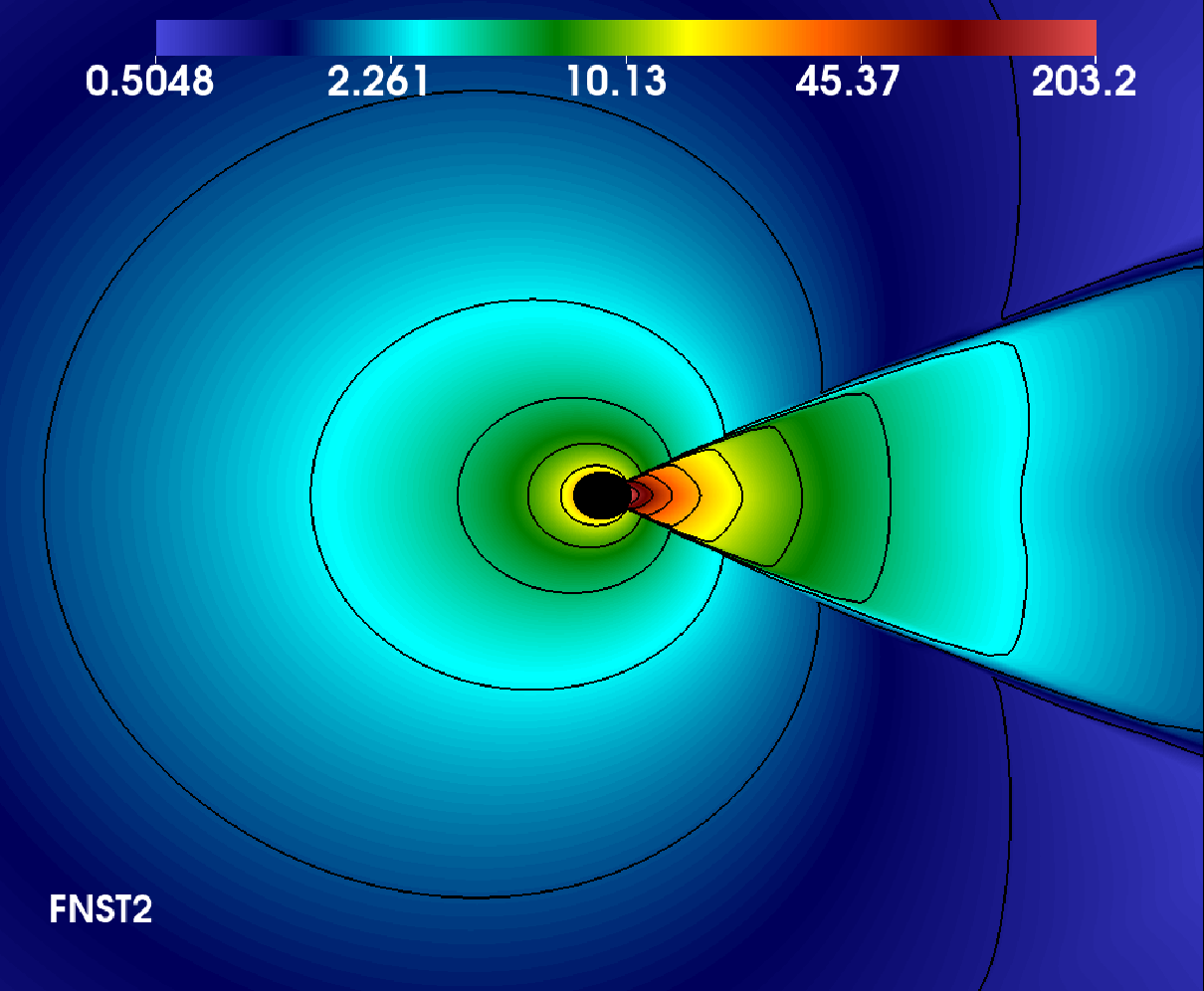}
\includegraphics[width=7.0cm,height=5.0cm]{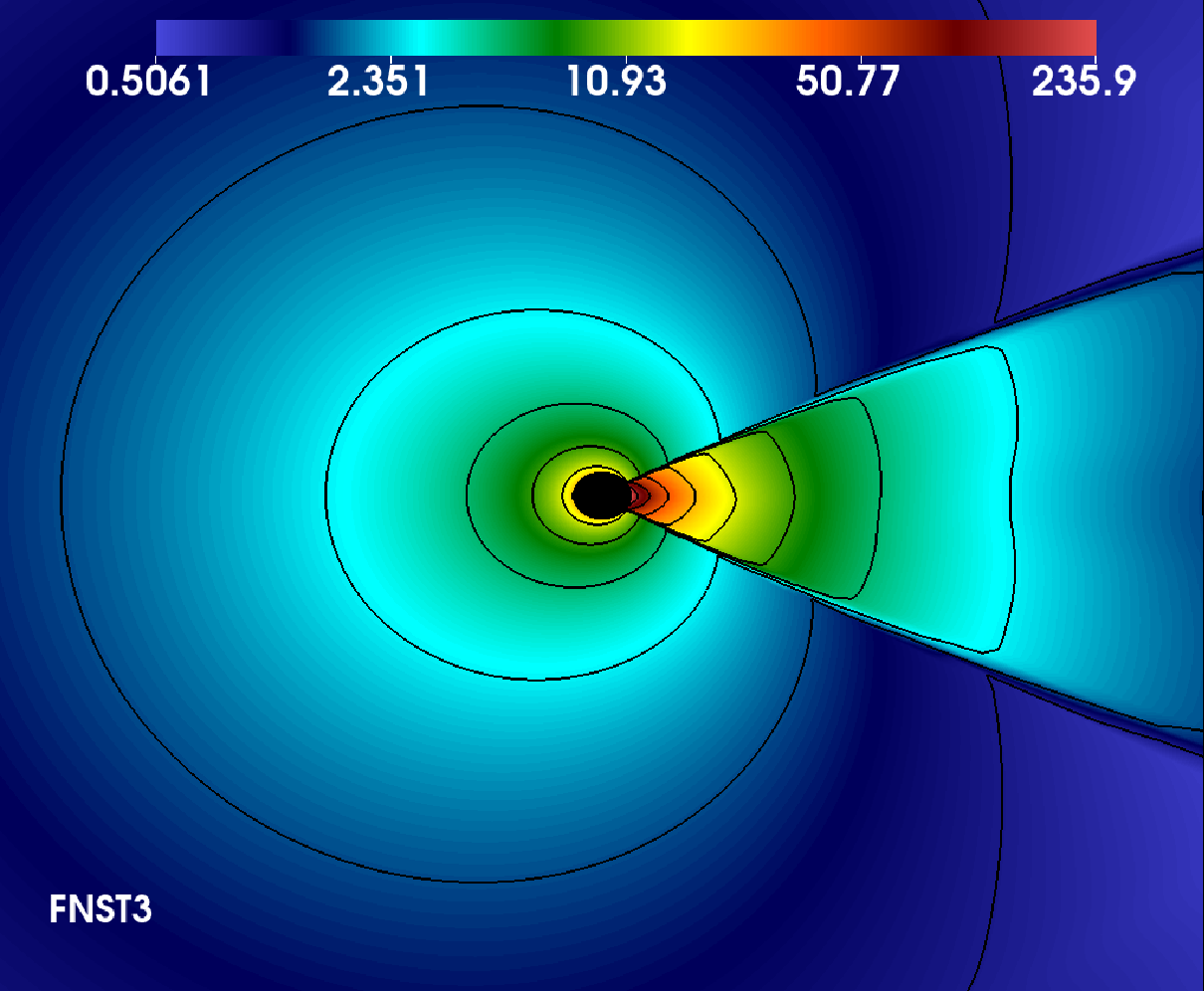}
\includegraphics[width=7.0cm,height=5.0cm]{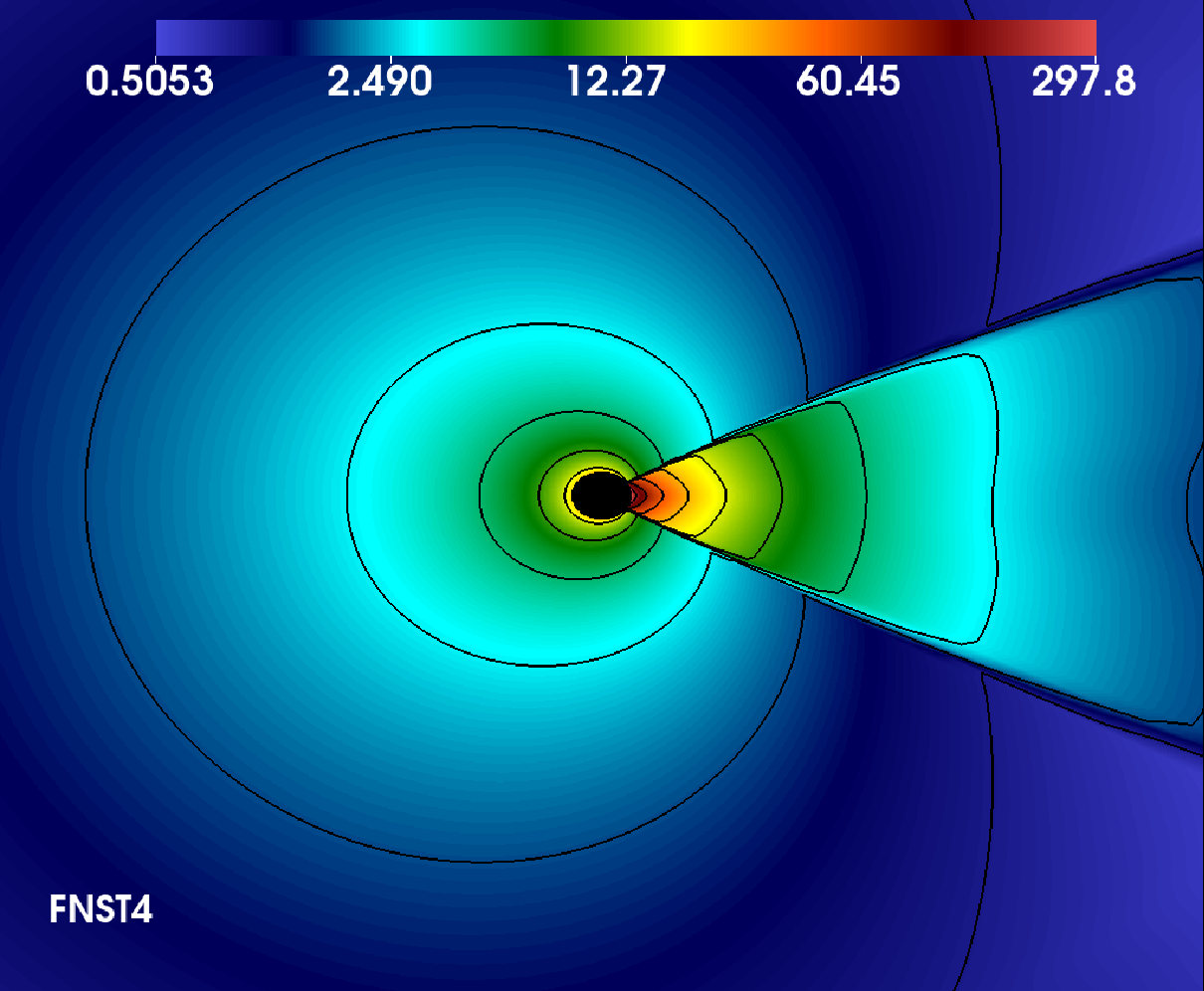}
\includegraphics[width=7.0cm,height=5.0cm]{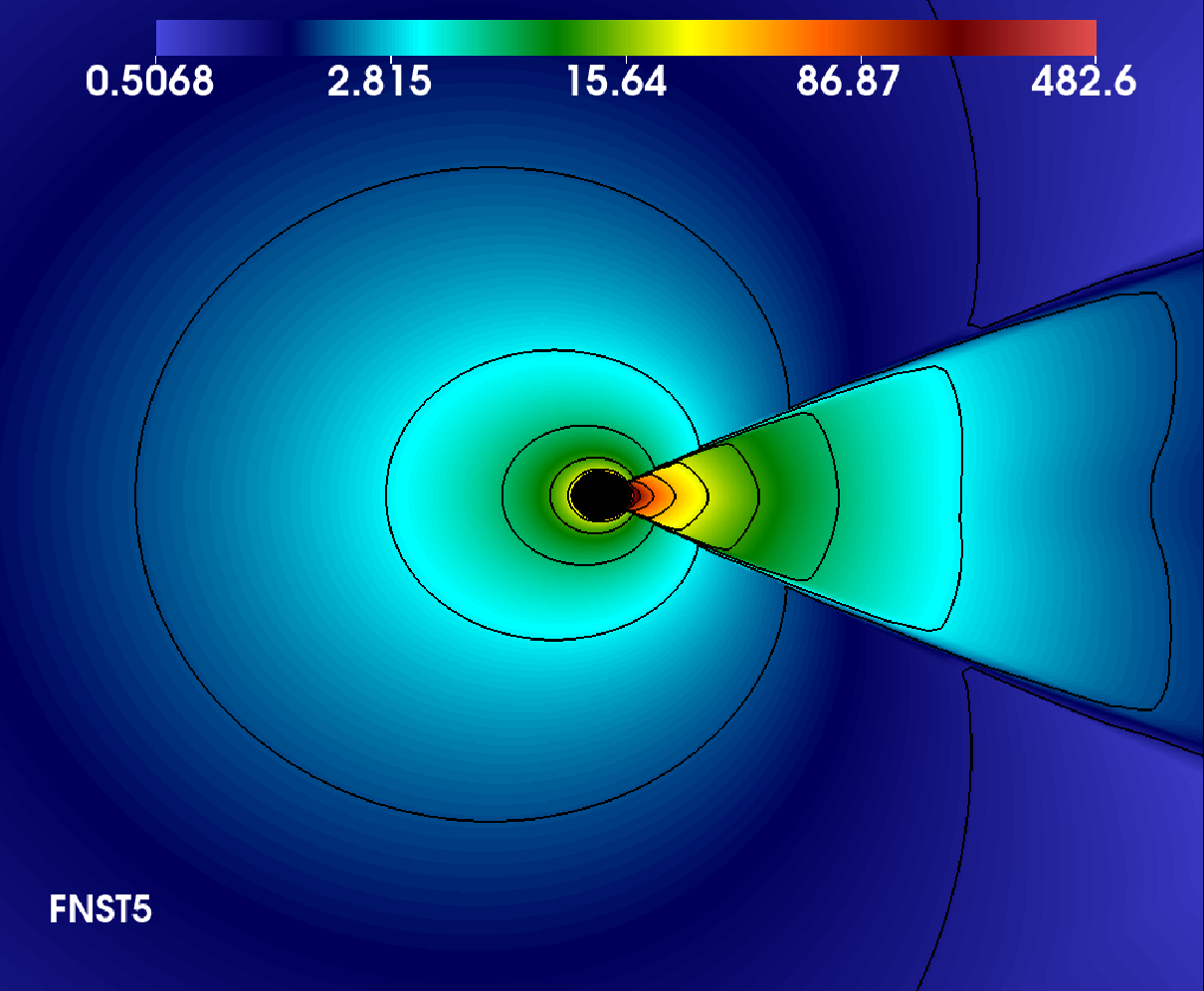}
\caption{The variation of the rest-mass density produced by BHL accretion in the two-dimensional equatorial plane is shown for the Schwarzschild and FNST gravity models. The computational view is zoomed to the region $-70M \leq x \leq 70M$ and $-70M \leq y \leq 70M$. The upper-left panel shows the shock-cone morphology formed in the Schwarzschild black-hole case, while the remaining panels display the shock-cone structure around FNST naked-singularity compact spacetimes as the deformation parameter $n$ decreases. The density distribution is shown by both the color map and the contour lines, allowing us to clearly identify the increase in the amount of matter accumulated inside the shock cone, particularly in the region close to the central compact object.}\label{2D_color}
\end{figure*}

\begin{figure*}
\centering
\includegraphics[width=12.5cm,height=7.0cm]{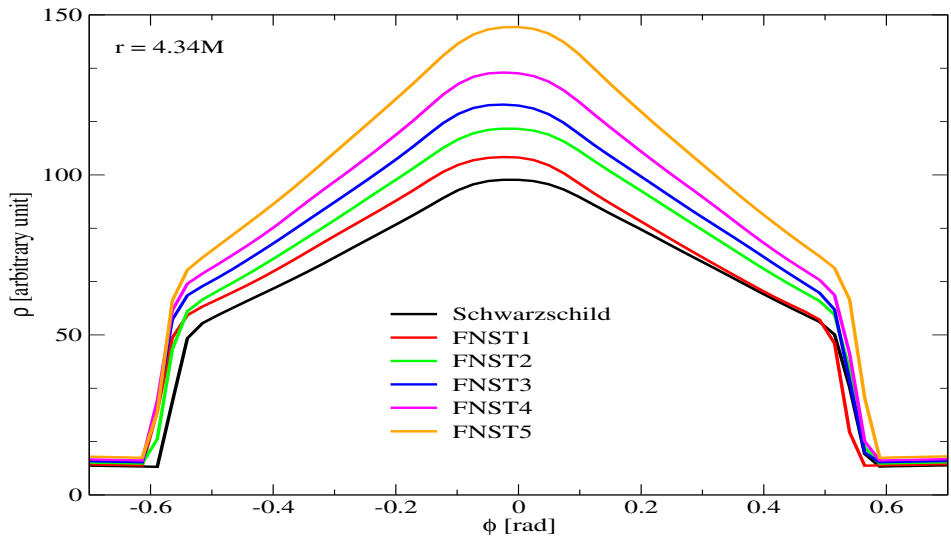}
\caption{At the fixed radial position $r=4.34M$, which is located outside the singular surface $r_s=2M/n$ for all FNST models considered here, the azimuthal variation of the rest-mass density is shown for the Schwarzschild and FNST models. The density reaches its maximum value around $\phi=0$, which corresponds to the central axis of the shock cone in the downstream region. As one moves away from this axis in both azimuthal directions, the density systematically decreases toward the edges of the cone. The increasing peak density from the Schwarzschild model to FNST5 shows that stronger FNST deformation enhances the accretion of matter inside the shock cone.}\label{den_1D_cut}
\end{figure*}

\begin{figure*}
\centering
\includegraphics[width=12.5cm,height=7.0cm]{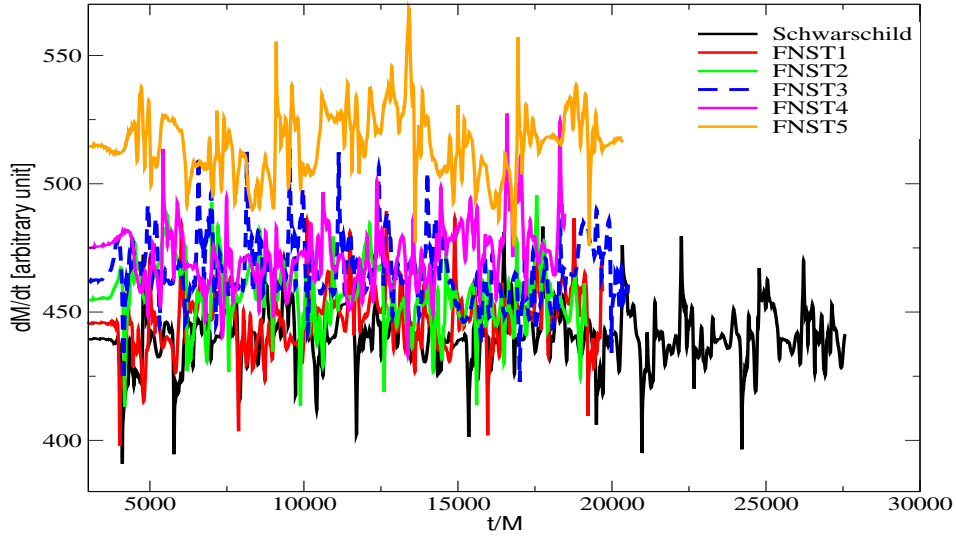}
\caption{The time evolution of the rest-mass flux through the inner excision boundary is shown at $r_{in}=3.4M$ for the Schwarzschild black-hole reference case and the FNST compact-spacetime models. The oscillatory behavior observed in the measured mass flux reflects the time-dependent dynamical structure of the shock cone and the modulation of matter crossing the inner radial excision boundary. Compared with the Schwarzschild case, the FNST models show a systematic increase in both the average inner-boundary mass flux and its temporal variability as the deformation parameter n decreases.}\label{accret}
\end{figure*}
\begin{figure*}
\centering
\includegraphics[width=7.0cm,height=5.0cm]{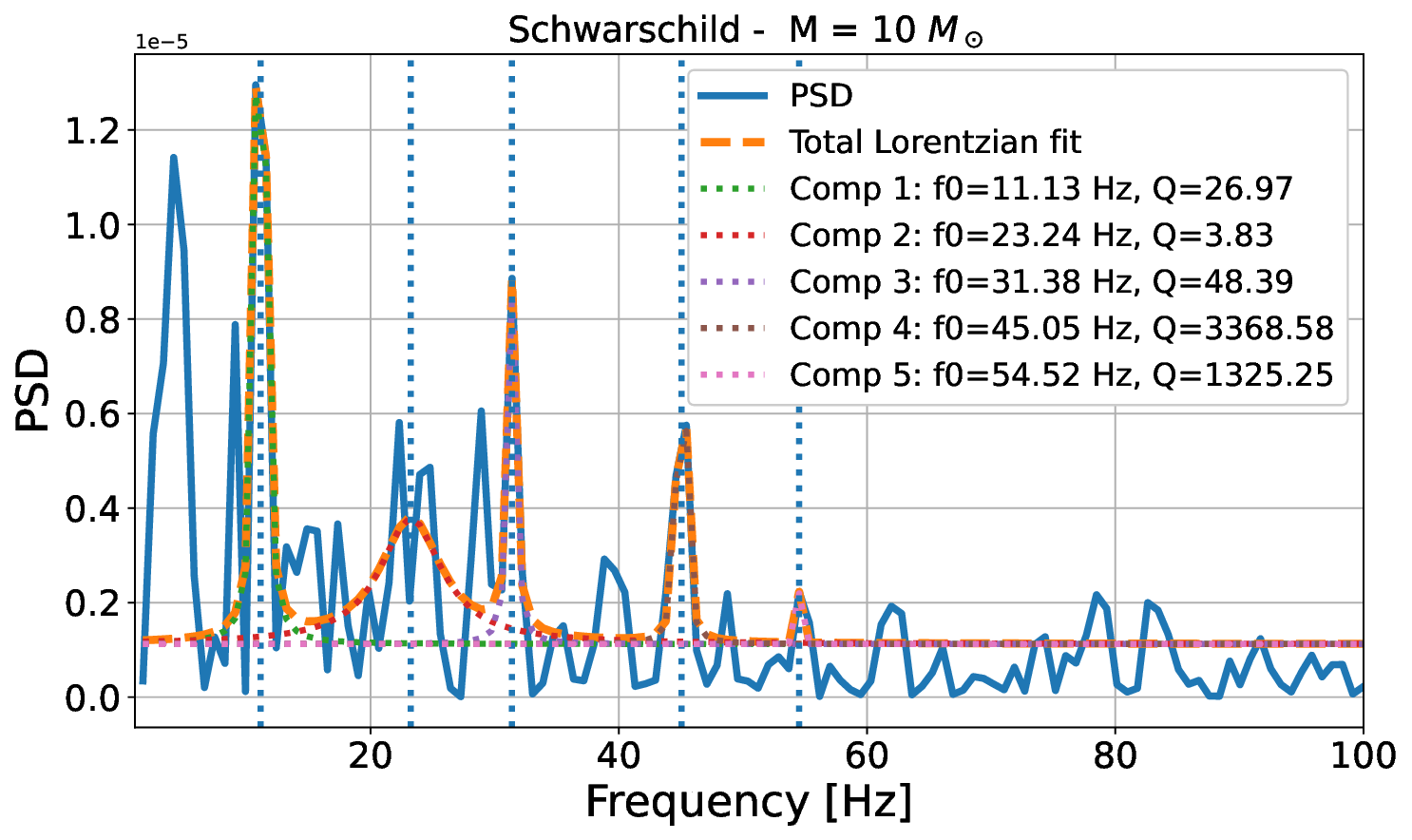}
\includegraphics[width=7.0cm,height=5.0cm]{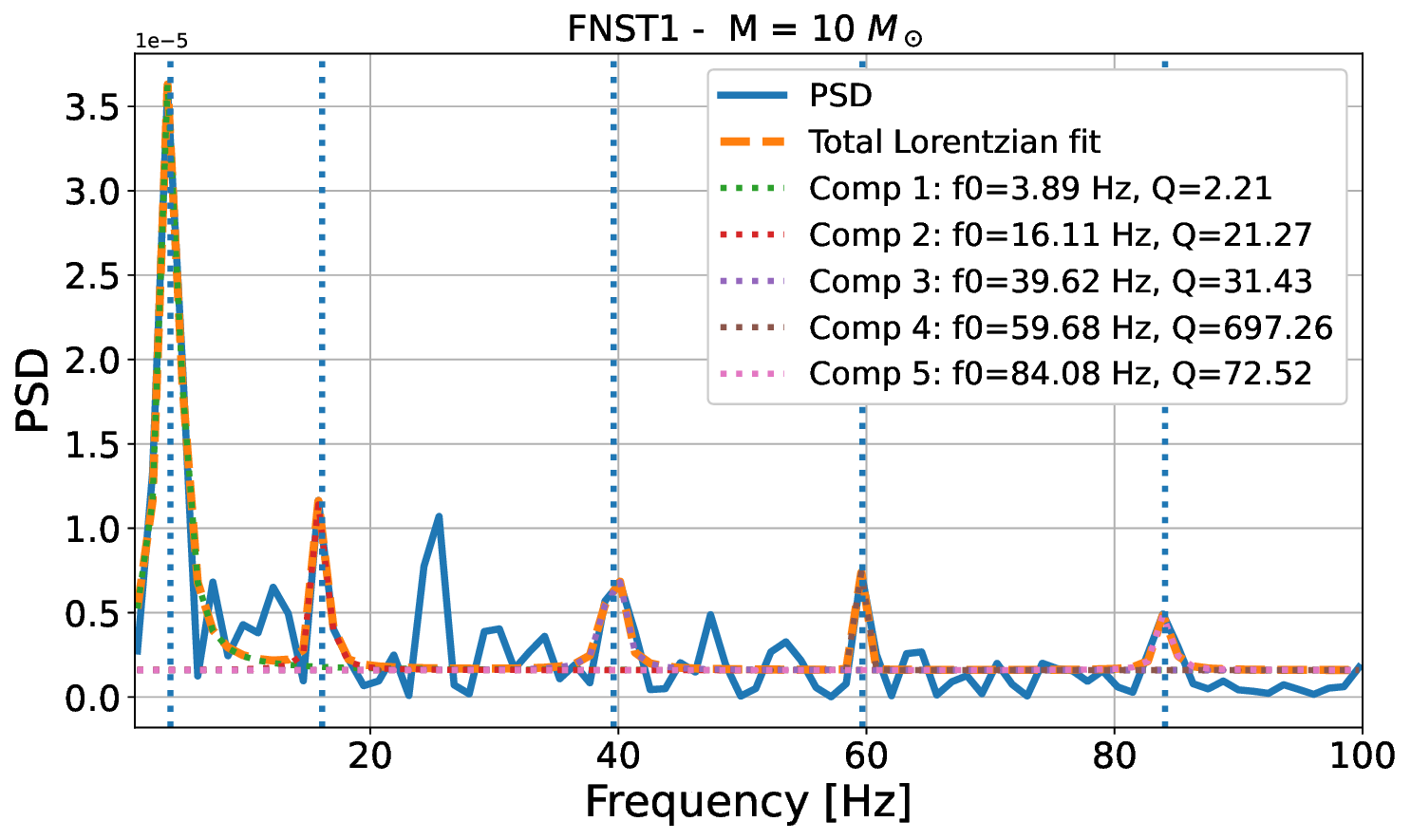}
\includegraphics[width=7.0cm,height=5.0cm]{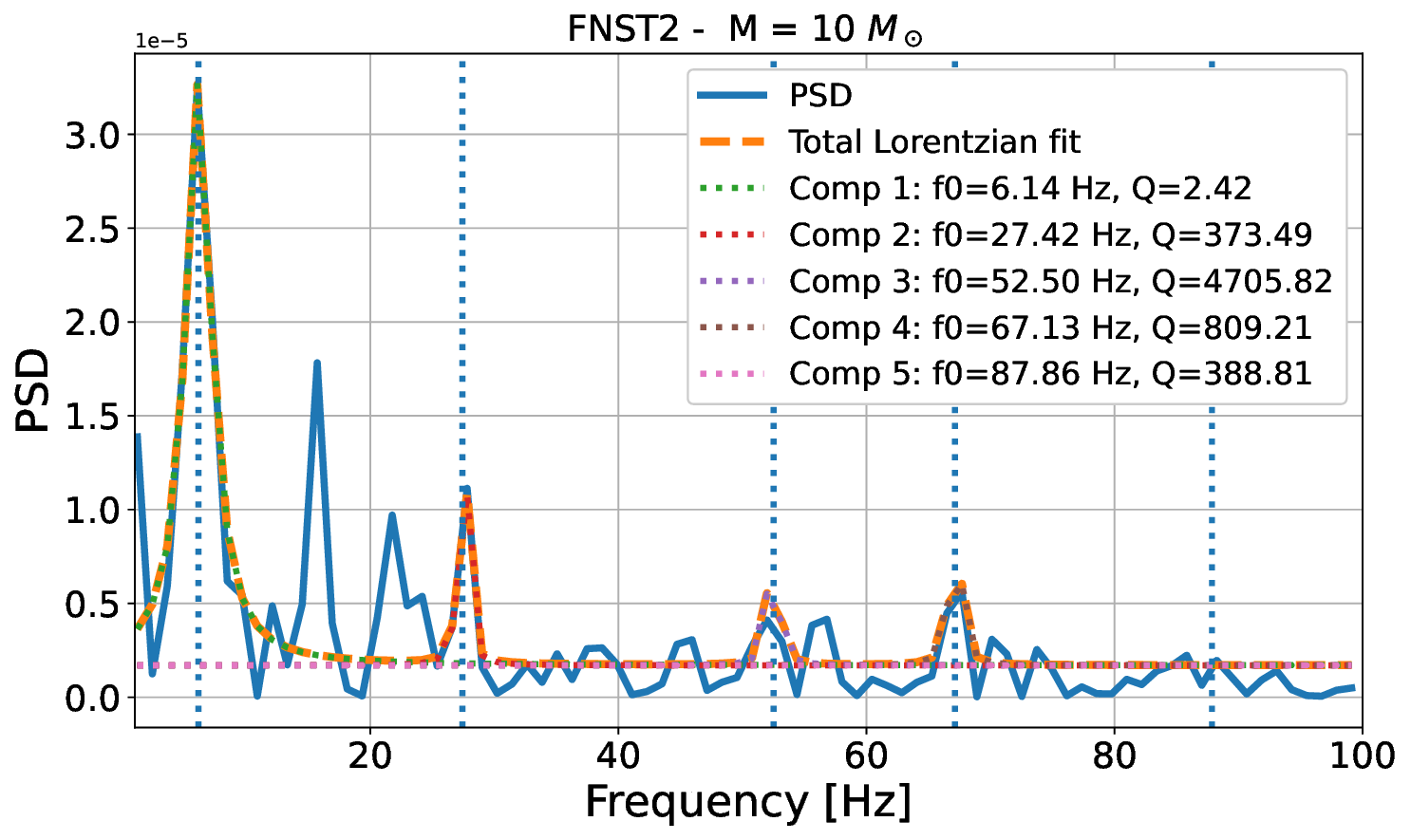}
\includegraphics[width=7.0cm,height=5.0cm]{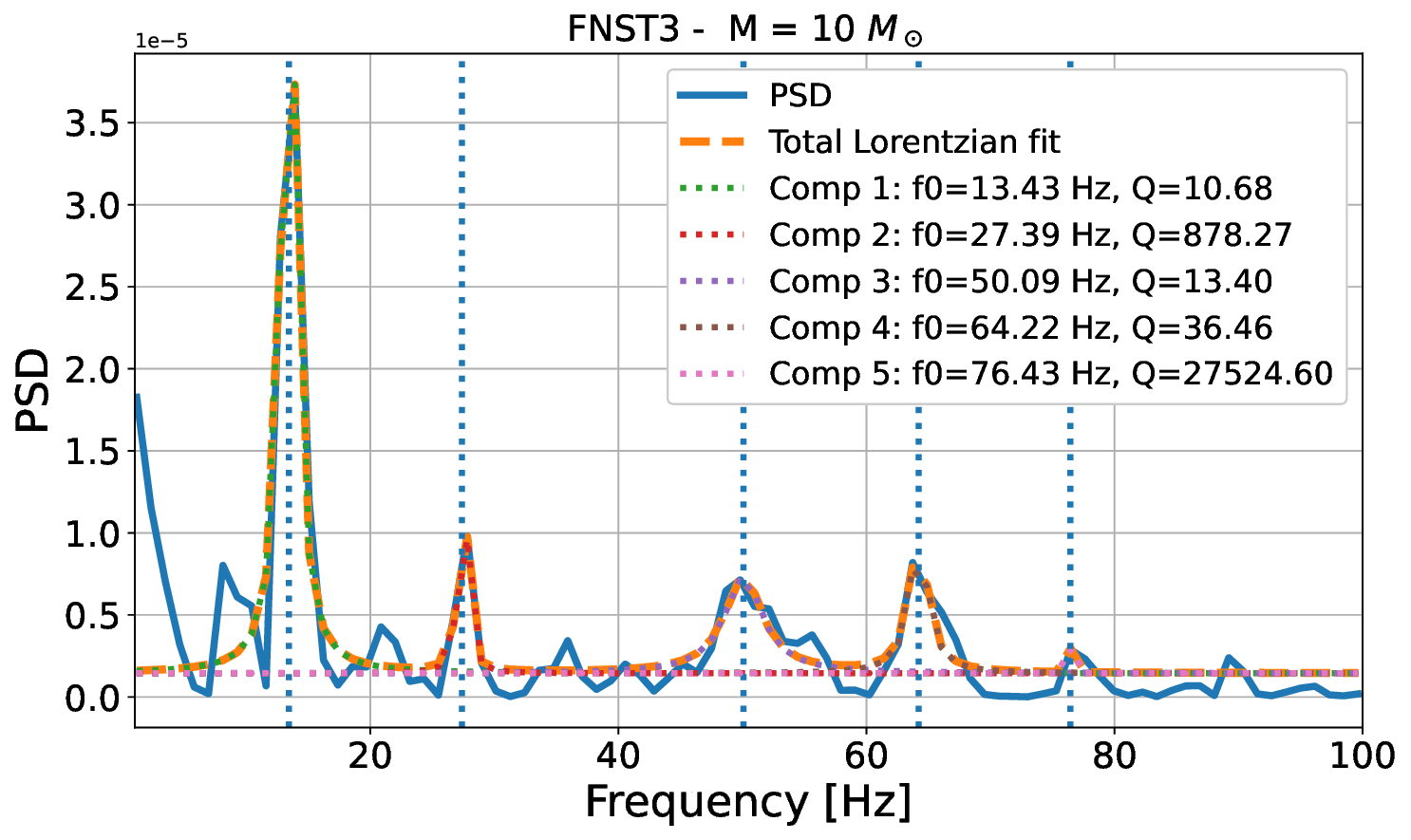}
\includegraphics[width=7.0cm,height=5.0cm]{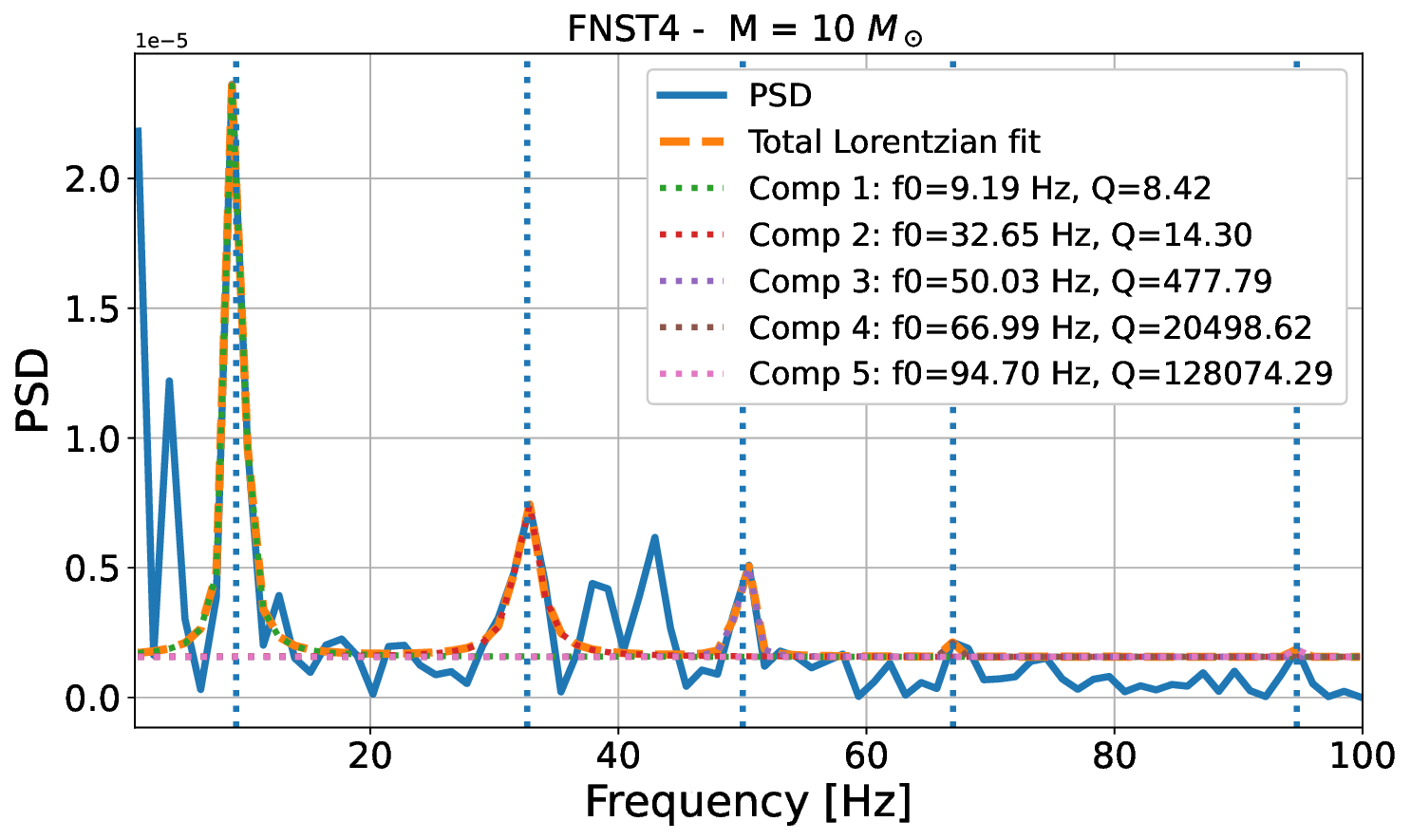}
\includegraphics[width=7.0cm,height=5.0cm]{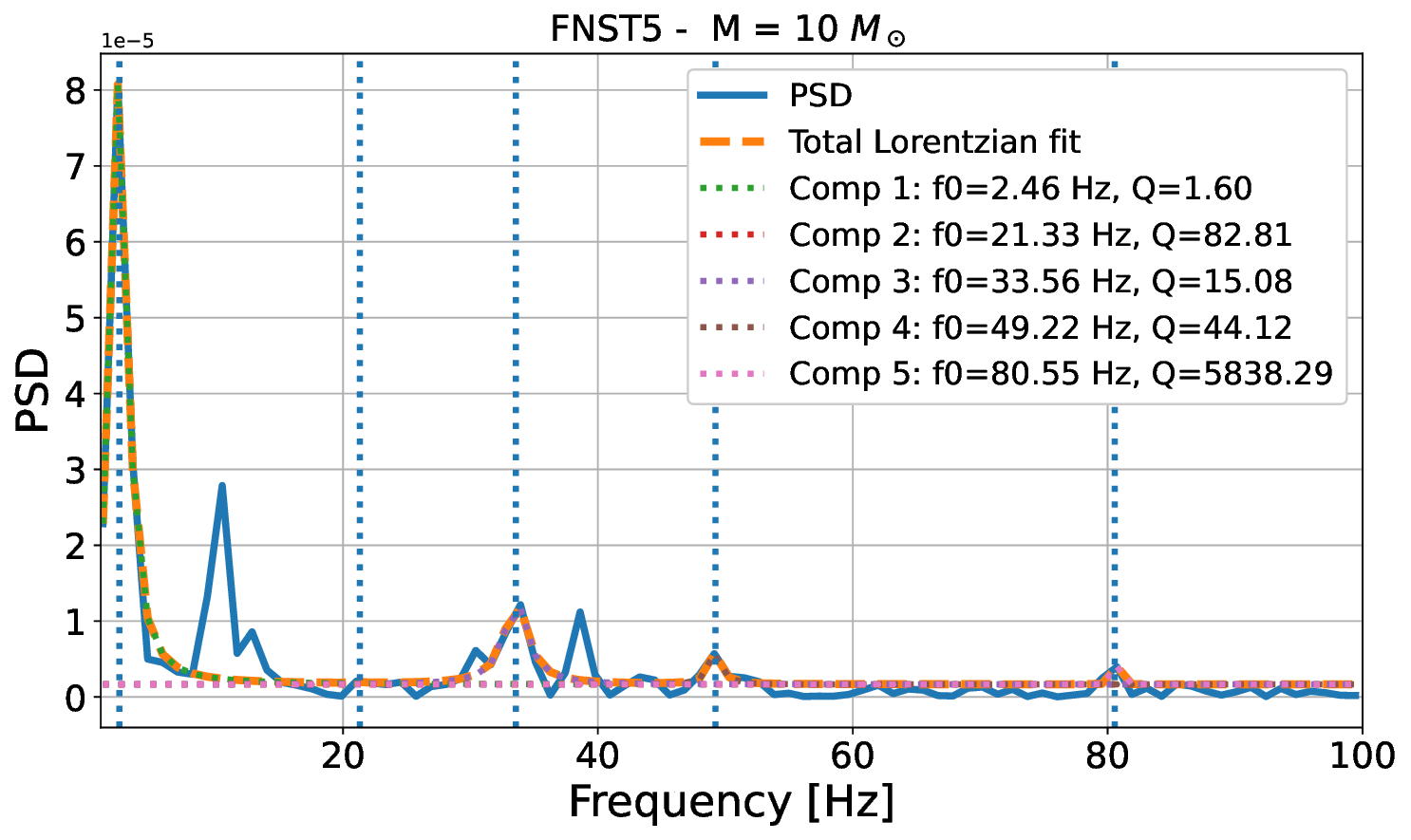}
\caption{PSD analysis computed from the rest-mass flux measured at the inner excision boundary $r_{in}=3.4M$ around the Schwarzschild black-hole reference case and the FNST compact-spacetime models. The blue curves show the numerically calculated PSDs, while the orange dashed curves represent the Lorentzian fits. The vertical dotted lines indicate the central frequencies of the fitted oscillatory components. From the upper panels to the lower panels, the decrease in the deformation parameter $n$ shows how strongly the solution deviates from the Schwarzschild case and how the properties and spectral appearance of the resulting QPO-like features are reorganized.}\label{QPOs}
\end{figure*}

\section{Hydrodynamical Accretion Dynamics in FNST Gravity}
\label{Num1}

The hydrodynamical evolution of the accreting plasma is investigated by solving the GRH equations on a fixed FNST gravity background. The numerical evolution is performed using a high-resolution shock-capturing scheme, which allows us to accurately resolve the discontinuities and shock fronts that develop along the boundaries of the shock cone around the central compact object \cite{Donmez2004ASS, Donmez2006AMC}. The matter is modeled as a perfect fluid, and the pressure is closed through an ideal-gas equation of state. In order to describe the physical mechanism by which matter is driven toward the central compact object, we adopt BHL accretion. In this configuration, matter is injected from the upstream region and is gravitationally focused toward the compact object, leading to the formation of a shock cone in the downstream region \cite{Donmez2012MNRAS, Donmez2024JCAP}. The simulations are carried out on the two-dimensional equatorial plane, which enables us to examine in detail the dynamical structure of the shock cone formed behind the central compact object. In addition, we show that the oscillatory behavior of the shock cone and the density variation of the matter trapped inside the cone depend sensitively on the spacetime parameters.

For the FNST models with $0<n<1$, the inner boundary condition should not be interpreted as a black-hole horizon condition. Since these spacetimes do not possess an event horizon, the surface $r_{\rm in}=3.4M$ is introduced as a numerical excision surface located outside the curvature-singularity surface $r_s=2M/n$. The purpose of this boundary is to remove matter that has already entered the unresolved innermost strong-field region, while preventing artificial reflections from propagating back into the computational domain. This treatment is therefore an absorbing sink prescription for the exterior hydrodynamical flow, not a claim that the matter crosses a regular event horizon. The shock-cone morphology, density enhancement, mass-flux variability, and PSD features discussed below are consequently interpreted as properties of the exterior accretion flow outside $r_{\rm in}$. In this sense, the computed diagnostic measures the rest-mass flux through the inner excision boundary, rather than the physical horizon accretion rate of a black hole. This distinction is important for horizonless or naked-singularity geometries, but it does not invalidate the study of the exterior BHL shock dynamics, because the shock cone and its oscillatory modes are generated in the resolved region outside the excision surface. A fully self-consistent treatment of the central singular region would require an additional physical model for the central object or for the high-curvature interior, which is beyond the scope of the present work. The present simulations should therefore be regarded as an exterior-flow study of FNST compact spacetimes, with the inner boundary acting as a controlled absorbing sink.

Outflow boundary conditions are imposed at the outer boundaries, except for the upstream injection region. At the inner radial boundary, located at $r_{\rm in}=3.4M$, we impose an absorbing outflow condition. For the Schwarzschild reference case, this boundary lies outside the event horizon and plays the usual role of an inner accretion boundary. For the FNST models with $0<n<1$, however, it is an excision boundary located outside the singular surface $r_s=2M/n$. Therefore, matter crossing $r_{\rm in}$ is removed from the computational domain, and no physical reflection from the unresolved central region is imposed. This choice allows us to focus on the exterior shock-cone dynamics while minimizing artificial boundary effects \cite{Donmez2024MPLA}.

The spacetime parameters used in the simulation models are given in Table~I. As throughout the paper, we set $M=1$ in geometrized units. The FN deformation parameter $n$ controls the deviation of the spacetime metric from the Schwarzschild geometry. For $n=1$, the metric reduces to the Schwarzschild metric, whereas smaller values of $n$ represent stronger scalar-field-induced deviations from GTR. For \(n=1\), the metric reduces to the Schwarzschild black-hole spacetime and the surface \(r_s=2M\) is the regular event horizon. For \(0<n<1\), however, the surface $r_s=\frac{2M}{n}$ is not an event horizon but a JNW-type curvature-singularity surface. Therefore, the FNST1--FNST5 models are interpreted as naked-singularity or horizonless compact-object spacetimes. In the FNST5 model, where $n=0.65$, this singular surface is located at $r_s=3.0769M$. To avoid the singular region and possible numerical artifacts, the inner radial boundary is chosen as $r_{\rm in}=3.4M$ for all models, while the outer radial boundary is fixed at $r_{\rm out}=100M$. The boundary $r_{\rm in}$ is therefore an inner excision boundary, not a black-hole horizon.

\begin{table}[ht]
\centering
\caption{Compact set of FNST gravity models used. Here, \(M=1\) is adopted in geometrized units. For \(n=1\), \(r_s=2M\) corresponds to the Schwarzschild event horizon. For \(0<n<1\), \(r_s=2M/n\) denotes the JNW-type curvature-singularity surface and not a regular black-hole horizon. The parameter $q$ is kept free because it does not appear explicitly in the metric coefficients of the spacetime written in Eq.~\ref{JNW}, but rather belongs to the underlying scalar-gravity coupling sector. Thus, the present model classification is based only on the metric deformation parameter $n$.}
\setlength{\tabcolsep}{12pt}

\begin{tabular}{ccc}
\hline
Model & $n$ & $r_s/M=2/n$ \\
\hline
FNST0 (Schwarschild) & 1.00 & 2.0000 \\
FNST1 & 0.90 & 2.2222 \\
FNST2 & 0.80 & 2.5000 \\
FNST3 & 0.75 & 2.6667 \\
FNST4 & 0.70 & 2.8571 \\
FNST5 & 0.65 & 3.0769 \\
\hline
\end{tabular}
\label{tab:fnst_models_compact}
\end{table}

\noindent
Since the metric coefficients used in the hydrodynamical simulations depend explicitly on the deformation parameter $n$ but not on $q$, the present numerical analysis should be interpreted as probing the exterior hydrodynamics of the FNST/JNW-type metric deformation controlled by $n$, rather than independently constraining the coupling parameter $q$.
 
\subsection{Shock Cone Morphology and Azimuthal Density Structure in FNST Gravity}
\label{Num2}
 
Fig.\ref{2D_color} shows the variation of the shock cone morphology formed by BHL accretion on the 2-dimensional equatorial plane as a function of the FN deformation parameter $n$. In all models, matter is injected from the left-hand side, which corresponds to the upstream region, and falls supersonically toward the central compact object. Due to gravitational focusing, the matter approaching the central compact object is strongly compressed. If the accreting material has sufficient angular momentum, it does not immediately cross the inner excision boundary. Instead, it is redirected and accumulated behind the compact object, leading to the formation of a shock cone on the right-hand side, namely in the downstream region. Each snapshot in Fig.\ref{2D_color} represents the state of the system long after it has reached a quasi-steady configuration. The color maps and contour lines in the snapshots show the variation of the rest-mass density. Therefore, the figure allows us to examine how the density of the plasma trapped inside the shock cone, as well as the surrounding accreting material outside the cone, changes across different spacetime models.  The upper-left panel shows the Schwarzschild solution, while the remaining panels correspond to the FNST models. These models demonstrate how the shock cone morphology changes as the deformation parameter $n$ moves away from its Schwarzschild value, $n=1$, indicating an increasing deviation from the Schwarzschild spacetime.

For the Schwarzschild model displayed in the upper-left panel of Fig. \ref{2D_color}, the accreting matter becomes highly collimated in the downstream region of the black hole, producing an approximately symmetric cone centered around $\phi$ is equal to zero. When the FNST deformation is included in the calculations, the overall shock cone structure is preserved. This indicates that the gravitational focusing effect remains efficient even in the scalar tensor gravity background. On the other hand, the morphology of the resulting cone changes systematically with decreasing values of the deformation parameter $n$. For the FNST1 and FNST2 models, which mildly deviate from the Schwarzschild spacetime, the shock cone almost preserves its original width. However, even in these models, the density tends to increase in the region close to the central compact object, where the gravitational field is stronger. In particular, for the FNST3, FNST4, and FNST5 models, which correspond to stronger deviations from the Schwarzschild solution, the width of the cone becomes slightly larger. At the same time, the density near the inner excision region increases significantly in these models. The density contours also become more extended, indicating that the scalar-field-induced deformation modifies the gravitational focusing of the incoming plasma.

The main role of the FN deformation parameter $n$ is to control the deviation from the Schwarzschild accretion morphology. Since the case $n=1$ corresponds to the Schwarzschild limit, a decrease in $n$ implies a stronger influence of the scalar field on the spacetime geometry. As $n$ decreases, the characteristic singular surface, defined as $r_s=2M/n$, moves outward. Thus, for the FNST models with $0<n<1$, the innermost strong-field region accessible to the flow is shifted to larger radii. As a result, the region close to the inner excision boundary is more strongly affected by the modified geometry. This behavior leads to a slight increase in the opening angle of the shock cone formed in the downstream region. At the same time, more matter accumulates near the compact object, which increases the density in this region and produces a more extended high-density wake behind the compact object. These results show that the morphology of the shock cone depends sensitively on the FNST spacetime parameter. Therefore, the shock cone structure can be used as a hydrodynamical diagnostic for distinguishing scalar tensor gravity models from the Schwarzschild solution.

Fig.\ref{den_1D_cut} shows the azimuthal variation of the rest-mass density at a fixed radial position, $r=4.34M$, for the Schwarzschild and FNST gravity models. This radius is located outside the singular surface for all FNST models considered here, where the gravitational field is strong. From these profiles, one can identify the location of the maximum density inside the shock cone formed in the downstream region, which occurs around $\phi=0$. At the same time, the angular opening of the shock cone can be inferred from the shock location along the azimuthal direction. The figure also shows that a low density plasma structure forms outside the shock cone, surrounding the central compact object. As one moves away from $\phi=0$ toward both positive and negative azimuthal directions, the density decreases in an almost symmetric manner. The azimuthal edges of the cone occur approximately at $|\phi| \simeq 0.55$-$0.60$ rad. At these locations, the density drops sharply to its minimum possible value. This sharp decrease indicates the transition from the dense shocked region inside the cone to the low density external flow outside the cone.

\subsection{Inner-Boundary Mass-Flux Oscillations and PSD Signatures in FNST Gravity}
\label{Num3}
The rest-mass flux is computed by measuring the flux of matter crossing the inner radial excision boundary located at $r_{\rm in}=3.4M$, outside the singular surface $r_s=2M/n$ for all FNST models considered here. For the Schwarzschild reference case, this quantity can be interpreted in the usual sense as an inner accretion-rate diagnostic. For the FNST models with $0<n<1$, however, it should be interpreted operationally as the mass flux through the absorbing excision boundary, not as a physical horizon accretion rate. This quantity provides one of the most direct diagnostics of the dynamical behavior of the exterior accretion flow, because it connects the hydrodynamical structure of the shock cone with the amount of matter transported toward the unresolved central region. Variations in the measured mass flux indicate that the shock cone undergoes oscillatory motion, while the trapped plasma experiences compression and rarefaction processes together with time-dependent instabilities. These oscillations are particularly important because variations in the inner-boundary mass flux can modulate the radiation emitted from the inner accretion region. Therefore, the temporal changes in this mass-flux signal provide a bridge between numerical simulations and observed astrophysical variability, giving useful clues for understanding luminosity variations and QPO-like features observed in X-ray binaries and other accreting compact-object systems.

Fig.\ref{accret} shows the time evolution of the rest-mass flux through the inner excision boundary for the Schwarzschild reference model and the FNST gravity models. The Schwarzschild model has the lowest average inner-boundary mass flux among all models. In this case, the measured mass flux oscillates around approximately $430$-$450$ in arbitrary units, and the oscillation amplitude remains more moderate compared with the FNST models. Intermittent sharp drops and peaks are also observed. This behavior is the result of the standard shock cone oscillations produced by BHL accretion in the Schwarzschild background. 
When the FNST deformation is introduced, the measured mass flux through the inner excision boundary increases systematically. In the FNST1 model, the inner-boundary mass flux remains close to the Schwarzschild value. However, its average value becomes slightly larger and the oscillation amplitude becomes somewhat stronger. The FNST2 and FNST3 models show a clear increase in both the mean inner-boundary mass flux and the variability amplitude. This indicates that the modified geometry enhances the amount of matter moving toward the central compact object and crossing the inner excision boundary. The FNST4 model continues the same behavior with a higher baseline accretion level. Finally, the FNST5 model exhibits the strongest inward mass transport, with maximum peaks in the measured mass flux reaching approximately $560$-$570$ in arbitrary units. Therefore, Fig.\ref{accret} shows that decreasing the deformation parameter $n$ increases the amount of matter crossing the inner excision boundary and strengthens the dynamical activity of the accretion flow.

When the FNST models given in Fig.\ref{accret} are compared with each other, the sequence from FNST1 to FNST5 shows a clear parameter-dependent trend. As the value of $n$ gradually decreases from the Schwarzschild value $n=1$, the characteristic singular surface $r_s=2M/n$ expands outward. Therefore, the inner strong-field region outside $r_s$ becomes more strongly affected by the scalar-field-induced deformation. This produces stronger gravitational focusing close to the central compact object. As a result, a larger amount of matter accumulates inside the shock cone, especially near the inner excision region, which leads to a systematic increase in the measured mass flux through the inner boundary. The FNST1 and FNST2 models correspond to mild deviations from Schwarzschild, while the FNST3 and FNST4 models represent intermediate to-strong deviations. The FNST5 model clearly differs from the others by producing the highest inner-boundary mass flux. In the models with stronger inner-boundary mass fluxes, the matter accumulated inside the shock cone around the compact object becomes denser.

The oscillations shown in Fig.\ref{accret} indicate that the accretion flow does not become completely steady even after the system reaches a quasi-stationary state. These oscillations are generally irregular, but they persist throughout the evolution. Physically, this behavior can be associated with the oscillatory motion of the shock-cone pattern near the compact object. The scalar tensor deformation significantly strengthens the hydrodynamical variability of the system. Compared with the Schwarzschild model, the FNST models produce stronger density accumulation and larger amplitudes in the measured inner-boundary mass flux. Therefore, larger deviations from Schwarzschild can generate more pronounced shock cone oscillations and quasi-periodic features in the observable accretion signal. In this sense, QPO-like signals of hydrodynamical origin may become observable. The time-domain fluctuations in in Fig.\ref{accret}  provide the input signal for the PSD analysis shown later in Fig.\ref {QPOs}. Thus, a strong FNST deformation not only increases the mass flux through the inner excision boundary but also modifies the variability behavior of the exterior flow. This means that the deformation parameter $n$ can leave observable imprints both on the amplitude of the inner-boundary mass-flux signal and on the frequency-domain structure of the emitted signal.

 Fig.\ref{QPOs} shows the PSD analyses computed from the inner-boundary mass-flux signals presented in Fig. \ref{accret} for the different models. These PSD studies reveal the parameter-dependent changes in the mode structure and characteristic frequencies of the perturbations trapped inside the shock cone around the central compact object, measured at the inner excision boundary $r=3.4M$, where the gravitational field is strong. The blue curve represents PSD, while the orange dashed curve shows the total Lorentzian fit. The vertical dotted lines indicate the central frequencies of the fitted Lorentzian components. The PSD analyses given in Fig. \ref{QPOs} provide direct information about the oscillatory behavior of the shocked plasma, the time-dependent motion of the shock cone, and the modulation of the matter crossing the inner radial excision boundary. In the Schwarzschild model shown in the upper-left panel of plot \ref{QPOs}, the approximate frequencies obtained from the Lorentzian fit are located around $11$ Hz, $23$ Hz, $31$ Hz, $45$ Hz, and $54$ Hz. These frequencies can be classified as moderate low frequency QPOs and moderate high frequency QPOs. Except for the peak around $23$ Hz, the other components have relatively large quality factors, $Q$, indicating that these peaks correspond to narrow and coherent oscillatory features. This shows that even in the Schwarzschild background, BHL accretion can produce QPO-like variability through hydrodynamical oscillations in the shock cone region.

 When the FNST deformation is included in the numerical simulations, the PSD structure changes significantly from one model to another as seen in plot \ref{QPOs}. In the FNST1 model, a dominant low frequency component appears near $3.89$ Hz, together with additional components around $16$ Hz, $39$ Hz, $59$ Hz, and $84$ Hz. Although the first peak is strong, it is relatively broad. In contrast, the higher frequency components are narrower and therefore more coherent. In the FNST2 model, the fitted frequency peaks appear around $6$ Hz, $27$ Hz, $52$ Hz, $67$ Hz, and $87$ Hz. In particular, the components around $27$ Hz, $52$ Hz, and $67$ Hz are important because they are relatively narrow and have high quality factors. This indicates the formation of more stable QPO-like signatures. In the FNST3 model, a strong peak appears around $13$ Hz, together with additional peaks around $27$ Hz, $50$ Hz, $64$ Hz, and $76$ Hz. The components around $27$ Hz and $76$ Hz are particularly coherent, suggesting that an intermediate deformation can excite long-lived oscillatory modes inside the shock cone. In the FNST4 case, peaks are formed around $9$ Hz, $32$ Hz, $50$ Hz, $67$ Hz, and $94$ Hz. Among all models, the highest-frequency component, around $94$ Hz, is observed in this model. In the FNST5 model, which corresponds to the strongest deformation, a very strong low frequency peak appears around $2.46$ Hz, together with additional peaks around $21$ Hz, $35$ Hz, $49$ Hz, and $80$ Hz. The lowest-frequency peak is strong but broad, whereas the peaks around $21$ Hz and $80$ Hz are much narrower and more coherent. This behavior shows that the deformation parameter $n$ does not simply increase all frequencies. Instead, by modifying the shock cone dynamics and the characteristic timescale of plasma motion, it reorganizes the oscillation spectrum.

 When the Schwarzschild model shown in Fig.\ref{QPOs} is compared with the FNST models, it becomes clear that the scalar tensor deformation leaves a distinct imprint on the frequency-domain behavior of the accretion flow. In the time domain, Fig.\ref{accret} shows that the FNST deformation increases both the average inner-boundary mass flux and the amplitude of its variability.  At the same time, Fig.\ref{QPOs} demonstrates that stronger variability produces a richer PSD structure with model-dependent Lorentzian components. Physically, the peaks in the PSD analysis can be interpreted as hydrodynamical modes associated with the oscillation of the shock cone. The compression and rarefaction cycles of the dense matter trapped inside the shock cone, together with the periodic or quasi-periodic variations of the mass flux near the inner excision boundary, can explain the origin of these oscillations. These modes are generally related to radial and azimuthal oscillation modes, in addition to the modes generated by the global oscillatory motion of the shock cone \cite{Donmez2024JCAP, Donmez2025JHEAp}. The peaks with high quality factors, $Q$, found in the numerical results are especially important because they correspond to narrow and coherent oscillations. Therefore, such peaks have a higher probability of being extracted as QPO-like features from observed light curves. However, a high $Q$ value alone is not sufficient. The peak must also have enough PSD power. Thus, the most observationally relevant peaks are those that are both strong in amplitude and narrow in frequency width.

Considering the strength of the peaks above the background noise and their coherence in Fig.\ref{QPOs}, the most prominent observable candidates are not necessarily only the peaks with the largest quality factor, $Q$. The most effective observable peak candidates are those that are both clearly above the PSD background and sufficiently narrow. In the Schwarzschild case, the components near $11$ Hz and $31$ Hz are the most reliable candidates. The peak near $11$ Hz has the strongest PSD power, while the component near $31$ Hz is narrower and more coherent. The high frequency peaks near $45$ Hz and $54$ Hz have very large $Q$ values; however, their observational relevance depends on how clearly their power remains above the local noise level. In the FNST1 model, the dominant peak near $3.89$ Hz is strong but broad. Therefore, it is better interpreted as a strong low frequency variability feature rather than a clean QPO. In this model, the more favorable coherent candidates are the peaks near $16$ Hz and $39$ Hz. On the other hand, the peaks near $59$ Hz and $84$ Hz appear as secondary candidates. These components are narrow, but their power is relatively weaker compared with the dominant features. In the FNST2 model, the peaks near $27$ Hz and $52$ Hz are the strongest candidates because they are clearly separated from the background and show high coherence. The peak near $67$ Hz can also be considered observationally relevant in this context. The component near $6$ Hz is strong, but it is broader and therefore less favorable as a clean QPO-like signal. In the FNST3 model, the peak near $13$ Hz is strong and observationally important. At the same time, the component near $27$ Hz is more coherent and therefore represents a stronger QPO-like candidate. The peak near $76$ Hz is highly coherent, but its detectability is expected to be weaker because its PSD power is relatively small. In the FNST4 model, the peaks near $32$ Hz and $50$ Hz are the most favorable candidates, since they combine stronger detectability with relatively narrow profiles. By contrast, the high frequency components near $67$ Hz and $94$ Hz have large $Q$ values, but their detectability may be more difficult because their power is closer to the background noise. In this model, the peak near $9$ Hz may also be one of the most favorable candidates. Although its $Q$ value is not as large as those of the higher frequency components, it has strong power and sufficient coherence, making it a suitable candidate for low frequency observations. In the FNST5 model, the peak near $2.46$ Hz is strong, but it represents broad low frequency variability. In contrast, the components near $21$ Hz and $80$ Hz are better coherent QPO-like candidates. The peaks near $35$ Hz and $49$ Hz in the same model can be regarded as moderate secondary features. Since the frequencies obtained here are computed for a mass parameter $M=10M_\odot$, they can be phenomenologically compared with stellar-mass compact-object or black-hole-candidate systems with similar mass estimates. If these results are compared with compact-object or black-hole-candidate systems with different mass estimates, intermediate mass BHs, or supermassive BHs, the corresponding frequencies can be estimated using the inverse-mass scaling relation.

\section{Hydrodynamical QPO-Like Modes within Observed Compact-Object Timing Ranges}\label{compare} 
The numerical results obtained provide a hydrodynamical interpretation for QPO-like variability within the frequency ranges observed in accreting compact-object and black-hole-candidate systems. In the numerical simulations, the shock cone formed by BHL accretion produces time-dependent oscillations in the inner-boundary mass-flux signal, and these oscillations appear as Lorentzian-like peaks in the PSD. For a mass parameter $10M_{\odot}$, the numerically extracted frequencies generally produce peaks from a few Hz up to approximately $100$ Hz. This numerical frequency range overlaps with the observed low-frequency and intermediate/high-frequency QPO phenomenology of stellar-mass black-hole candidates. In this sense, the model does not require a one-to-one identification with a specific observed peak, but it naturally produces QPO-like frequencies in the same astrophysically relevant range. For example, low frequency QPOs observed in BH binary systems generally fall in the range of $0.1$-$30$ Hz \cite{Rodriguez:2004kg, Zhang:2014dra}. On the other hand, timing features around $41$ Hz and $67$ Hz observed from the source GRS 1915+105 \cite{Belloni:2013qka, Belloni:2019sot, Sreehari:2020jge}, while high frequency QPOs near $184$ Hz and $276$ Hz reported from XTE J1550-564 \cite{Varniere:2018zea}. The Schwarzschild and FNST models discussed in Fig.\ref{QPOs} produce several peaks in the low to-intermediate frequency range. In addition, the stronger FNST cases generate higher frequency components that are compatible with the observed high frequency regime. Therefore, the shock cone oscillations whose dynamical structure is investigated in detail may provide a possible hydrodynamical physical mechanism for showing phenomenological overlap with the timing variability reported from sources such as GRS 1915+105.

A physically meaningful observational comparison can be established by analyzing the QPO-like modes that are hydrodynamically generated and 
extracted from the PSD of the inner-boundary mass-flux variability. Their astrophysical relevance can be evaluated by examining whether their characteristic frequencies, coherence properties, and spectral powers fall within the timing ranges reported for accreting BH systems. Among the peaks obtained from the numerical simulations, the components near $11$ Hz and $31$ Hz in the Schwarzschild model fall mainly within the LFQPO band. The dominant peaks in the FNST1 model lie in a similar frequency range. In particular, the broad component near $3.89$ Hz and the more coherent peaks near $16$ Hz and $39$ Hz make this model more suitable for explaining low frequency variability rather than sharp high frequency QPOs. The results obtained for the FNST2 model show a stronger agreement with the observed timing features. The numerically extracted components near 27 Hz, 52 Hz, and 67 Hz fall within the frequency range  $40$-$67$ Hz  of timing features reported for GRS 1915+105. In particular, the peaks near $52$ Hz and $67$ Hz are more important because they not only fall within the observed frequency range, but also display stronger coherence. The FNST3 model shows a similar consistency with observations. The QPO-like frequencies obtained near $50$ Hz, $64$ Hz, and $76$ Hz are compatible with the HFQPO range observed in GRS~1915+105. Therefore, among the models with moderate deformation, FNST2 and FNST3 show the closest phenomenological overlap with the GRS 1915+105-like timing range among the models considered here.

The strongly deformed FNST4 and FNST5 models provide a broader spectral coverage. Therefore, these models may be more effective for explaining observational sources that exhibit high frequency timing structures or richer peak distributions. In the FNST4 model, favorable peaks appear near $32$ Hz, $50$ Hz, and $67$ Hz, while an additional higher frequency component is also found near $94$ Hz. In this sense, FNST4 provides one of the strongest connections with the low , intermediate , and high frequency timing features observed from the source GRS~1915+105 \cite{Belloni:2019sot}. The FNST5 model shows a strong low frequency feature near $2.46$ Hz and produces more coherent peaks near $21$ Hz and $80$ Hz. Therefore, this model may better represent systems in which strong low frequency variability coexists with weaker higher frequency oscillations. On the other hand, the frequencies obtained for a BH mass of $M=10M_{\odot}$. For compact-object sources or black-hole candidates with different mass estimates, an inverse-mass scaling must be applied before making a direct comparison. After this scaling, high frequency QPOs such as the $184$ Hz and $276$ Hz components observed from XTE~J1550-564 \cite{Rodriguez:2004kg}, the $300$ Hz and $450$ Hz components observed from GRO~J1655-40 \cite{Mendez:2013yda}, or the $165$ Hz and $241$ Hz frequency pair reported from H1743-322 \cite{Homan:2004pp} may be brought into comparison with the present numerical results. Such a comparison may require considering a smaller effective BH mass for these sources, a different radial excitation region, stronger relativistic effects, or additional oscillation modes that are not fully captured in the present two-dimensional BHL simulations. Therefore, among the models studied in this paper, FNST2-FNST4 provide the closest phenomenological overlap with the GRS 1915+105-like QPO frequency range among the models considered here. In contrast, the Schwarzschild and FNST1 models mainly support low frequency variability, while the FNST5 model favors strong broad low frequency oscillations together with selected coherent higher frequency components.

We emphasize that the comparison with observed QPOs is phenomenological. The present simulations do not include a radiative transfer calculation, a physical emission model, or a direct mode-identification scheme. Therefore, the frequency comparison should not be interpreted as a unique explanation of the timing properties of any individual source. Instead, the purpose of the comparison is to show that the hydrodynamically generated inner-boundary mass-flux oscillations in the resolved exterior flow can produce QPO-like frequencies that overlap with the ranges reported for stellar-mass black-hole candidates. A more direct observational test would require additional modeling of the emission process, source parameters, and the mapping between hydrodynamical oscillation modes and observed light-curve variability.

\section{Conclusions}\label{Concl} 
We investigate the hydrodynamical properties of BHL accretion around a static and spherically symmetric compact-object solution in FNST gravity. The Schwarzschild limit $n=1$ corresponds to a regular black hole, while the deformed cases $0<n<1$ correspond to JNW-type naked-singularity geometries rather than regular black holes.  The simulations presented in this study are obtained by solving the GRH equations on a fixed FNST background. In order to accurately resolve the shock waves formed around the central compact object, numerical solutions are performed using a high-resolution shock-capturing scheme. The simulations are carried out by assuming that the accreting matter obeys an ideal-gas equation of state. By varying the deformation parameter $n$, which distinguishes the FNST spacetime from the Schwarzschild solution, we numerically show how the morphology of the shock cone formed around the central compact object, the density distribution of the accreting plasma, the time-dependent inner-boundary mass flux, and the PSD signatures computed from these signals deviate from the Schwarzschild limit. The numerical results show that the scalar-field-induced deformation does not destroy the shock cone mechanism formed around naked-singularity or horizonless compact-object geometries. However, it systematically and observationally modifies both the temporal and spatial behavior of the accretion flow.

The numerically obtained shock cone morphology and azimuthal density profiles show that the deformation parameter $n$ controls the strength of the deviation from the Schwarzschild accretion pattern. In the Schwarzschild case, the shock cone formed in the downstream region is narrow, well collimated, and almost symmetric around the downstream axis, namely around $\phi=0$. As n decreases, corresponding to a stronger FNST deformation of spacetime, the characteristic singular surface $r_s=2M/n$ moves outward, and the inner strong-field region outside this surface becomes more strongly affected by the modified geometry.  The azimuthal density profile taken at a fixed radius quantitatively confirms this behavior. The maximum density formed around the central axis of the shock cone, which represents the amount of matter accumulated inside the cone, systematically increases from the Schwarzschild model to FNST5. This confirms that stronger scalar tensor deformation significantly enhances the amount of matter trapped inside the shock cone.

Both the inner-boundary mass flux and the PSD signatures clearly show that the system remains dynamically active even after it reaches a steady state. The mass flux measured at the inner excision boundary of the computational domain,
$r_{in}=3.4M$, which lies outside the singular surface for all FNST models considered here, exhibits persistent oscillations due to the motion of the shock cone, the compression and rarefaction of the trapped plasma, and the modulation of the matter crossing the inner radial boundary. Compared with the Schwarzschild case, the FNST models show a systematic increase in both the inner-boundary mass flux and the amplitude of its variability as the deformation parameter $n$ decreases.
The PSD analysis computed from the inner-boundary mass-flux signal shows that the deformation parameter reorganizes the oscillation spectrum. Each FNST model produces a distinct set of Lorentzian components representing different physical states. These appear as model-dependent central frequencies, spectral powers, and quality factors. This demonstrates that scalar-tensor deformation modifies not only the amount of matter transported toward the central compact object and across the inner boundary, but also the frequency-domain structure of the accretion signal.

The comparison of the numerical results obtained in this paper with the observed timing ranges of black-hole candidates indicates that the extracted features are astrophysically relevant and may therefore provide a phenomenological signature of the underlying spacetime.  The numerical PSD peaks computed are likely to be hydrodynamically generated QPO-like modes, rather than frequencies imposed only by the geodesic motion of test particles. For sources modeled with mass parameter $M=10M_{\odot}$, most of the extracted frequencies appear in the range from a few Hz up to approximately $100$ Hz. This range overlaps with the observed LFQPO and intermediate/HFQPO frequency ranges of stellar-mass BHs. In the absence of deformation, or in the presence of only weak deformation, the Schwarzschild and FNST1 models mainly produce low frequency oscillations, such as the components near $11$ Hz, $31$ Hz, $3.89$ Hz, $16$ Hz, and $39$ Hz. Therefore, these models are more suitable for explaining broad LFQPO-like behavior rather than sharp high frequency QPO features.

Among the FNST models, FNST2, FNST3, and FNST4 show a strong consistency with the observed timing features reported from the source GRS~1915+105. In the FNST2 model, important peaks appear near $27$ Hz, $52$ Hz, and $67$ Hz, while the FNST3 model produces components near $50$ Hz, $64$ Hz, and $76$ Hz. In the FNST4 model, a broader spectral coverage is observed, with peaks near $32$ Hz, $50$ Hz, and $67$ Hz, together with an additional high frequency component near $94$ Hz. Therefore, the models with moderate-to-strong deformation appear as the most promising models showing the closest phenomenological overlap with GRS 1915+105-like QPO ranges through shock cone oscillations. On the other hand, the FNST5 model favors strong broad low frequency variability together with selected coherent higher frequency components near $21$ Hz and $80$ Hz. This makes it more suitable for systems in which strong low frequency variability coexists with weaker high frequency features. For compact-object sources with masses different from  $10M_{\odot}$, such as XTE J1550-564, GRO J1655-40, and H1743-322, as well as intermediate-mass and supermassive black-hole candidates, the corresponding frequencies should be compared after applying the inverse-mass scaling relation.

Finally, the present results show that the FN deformation parameter significantly modifies the morphological structure formed around the central compact object, its time-dependent evolution, and the spectral timing properties of the BHL accretion flow. In future work, we aim to extend this analysis by performing three-dimensional simulations and exploring different accretion configurations in order to understand how the hydrodynamical QPO-like modes are affected. Such an extension will also allow us to make the simulations more realistic for astrophysical systems and to investigate how the numerical results can explain the timing properties observed in black-hole-candidate sources.

\bibliography{references}

@article{yousaf2025implications,
  title={Implications of modified Gauss-Bonnet gravity on gravastar-like structures: High-energy stability and electromagnetic effects},
  author={Yousaf, M and Asad, H and Aslam, M.},
  journal={High Energy Density Phys.},
  volume={57},
  pages={101221},
  year={2025},
  publisher={Elsevier}
}

@article{Davlataliev:2026vkx,
    author = "Davlataliev, Akbar and Turimov, Bobur and Ahmedov, Bobomurat and Vyblyi, Yuri and Yuan, Chengxun and Zhou, Chen",
    title = "{Spherically-symmetrical vacuum solution in Freund-Nambu scalar-tensor gravity}",
    eprint = "2603.11500",
    archivePrefix = "arXiv",
    primaryClass = "gr-qc",
    month = "3",
    year = "2026"
}

@article{donmez2026relativistic,
  title={Relativistic accretion process onto rotating black holes in Einstein-Euler-Heisenberg nonlinear electrodynamic gravity},
  author={Donmez, Orhan and Mustafa, G and Chaudhary, Himanshu and Yousaf, M and Bouzenada, Abdelmalek and Ditta, Allah and Atamurotov, Farruh},
  journal={Phys. Dark Universe},
  volume={52},
  pages={102271},
  year={2026},
  publisher={Elsevier}
}

@article{donmez2026accretion,
  title={Accretion flow around Kerr metric in the infra-red limit of asymptotically safe gravity},
  author={Donmez, Orhan and Ghosh, Sushant G and Yousaf, Muhammad and Mustafa, Ghulam and Atamurotov, Farruh},
  journal={J. Cosmol. Astropart. Phys.},
  volume={2026},
  number={04},
  pages={045},
  year={2026},
  publisher={IOP Publishing}
}

@ARTICLE{belloni2012high,
       author = {{Belloni}, T.~M. and {Sanna}, A. and {M{\'e}ndez}, M.},
        title = "{High-frequency quasi-periodic oscillations in black hole binaries}",
      journal = {Mon. Not. Roy. Astron. Soc.},
         year = 2012,
        month = nov,
       volume = {426},
       number = {3},
        pages = {1701-1709},
          doi = {10.1111/j.1365-2966.2012.21634.x},
archivePrefix = {arXiv},
       eprint = {1207.2311},
 primaryClass = {astro-ph.HE},
       adsurl = {https://ui.adsabs.harvard.edu/abs/2012MNRAS.426.1701B}
}

@article{yousaf2024fuzzy,
  title={Fuzzy black hole models in {$f (\mathcal{G}) $} Gravity},
  author={Yousaf, M and Asad, H and Almutairi, Bander and Hasan, S and Khan, A S},
  journal={Phys. Scr.},
  volume={99},
  pages={115270},
  year={2024},
  publisher={IOP Publishing}
}

@article{singh2026probing,
  title={Probing strong-gravity chaos in rotating Kerr-Bertotti-Robinson black holes},
  author={Singh, Pradeep and Kala, Shubham and Nandan, Hemwati and Yousaf, M and Atamurotov, Farruh and Mustafa, G},
  journal={Chaos, Solitons \& Fractals},
  volume={208},
  pages={118379},
  year={2026},
  publisher={Elsevier}
}

@article{donmez2026disformal,
  title={Disformal Kerr Imprints on BHL Accretion: Shock Morphology, PSD Signatures, and Observational QPO Counterparts},
  author={Donmez, Orhan and Yousaf, M and Khan, Imtiaz and Mustafa, G},
  journal={arXiv preprint arXiv:2605.15686},
  year={2026}
}

@article{johannsen2013photon,
  title={Photon rings around Kerr and Kerr-like black holes},
  author={Johannsen, Tim},
  journal={Ast. J.},
  volume={777},
  number={2},
  pages={170},
  year={2013},
  publisher={The American Astronomical Society}
}

@article{Homan:2004pp,
    author = "Homan, Jeroen and Miller, Jon M. and Wijnands, Rudy and van der Klis, Michiel and Belloni, Tomaso and Steeghs, Danny and Lewin, Walter H. G.",
    title = "{High- and low-frequency quasiperiodic oscillations in the x-ray light curves of the black hole transient H1743-332}",
    eprint = "astro-ph/0406334",
    archivePrefix = "arXiv",
    doi = "10.1086/424994",
    journal = "Astrophys. J.",
    volume = "623",
    pages = "383--391",
    year = "2005"
}

@article{Rodriguez:2004kg,
    author = "Rodriguez, J. and Corbel, S. and Kalemci, E. and Tomsick, J. A. and Tagger, M.",
    title = "{An x-ray timing study of xte j1550-564: evolution of the low frequency qpo for the complete 2000 outburst}",
    eprint = "astro-ph/0405398",
    archivePrefix = "arXiv",
    doi = "10.1086/422672",
    journal = "Astrophys. J.",
    volume = "612",
    pages = "1018--1025",
    year = "2004"
}

@article{Mendez:2013yda,
    author = "Mendez, Mariano and Altamirano, Diego and Belloni, Tomaso and Sanna, Andrea",
    title = "{The phase lags of high-frequency quasi-periodic oscillations in four black-hole candidates}",
    eprint = "1308.0142",
    archivePrefix = "arXiv",
    primaryClass = "astro-ph.HE",
    doi = "10.1093/mnras/stt1431",
    journal = "Mon. Not. Roy. Astron. Soc.",
    volume = "435",
    pages = "2132",
    year = "2013"
}

@article{Zhang:2014dra,
    author = "Zhang, Wenda and Yu, Wenfei",
    title = "{Low frequency QPOs and possible change in the accretion geometry during the outbursts of Aquila X$-$1}",
    eprint = "1412.0960",
    archivePrefix = "arXiv",
    primaryClass = "astro-ph.HE",
    doi = "10.1088/0004-637X/805/2/139",
    journal = "Astrophys. J.",
    volume = "805",
    number = "2",
    pages = "139",
    year = "2015"
}

@article{Varniere:2018zea,
    author = "Varniere, Peggy and Rodriguez, Jerome",
    title = "{Looking for the Elusive 3:2 Ratio of High-frequency Quasi-periodic Oscillations in the Microquasar XTE J1550{\ensuremath{-}}564}",
    eprint = "1808.06823",
    archivePrefix = "arXiv",
    primaryClass = "astro-ph.HE",
    doi = "10.3847/1538-4357/aad774",
    journal = "Astrophys. J.",
    volume = "865",
    number = "2",
    pages = "113",
    year = "2018"
}

@article{Sreehari:2020jge,
    author = "Sreehari, H. and Nandi, Anuj and Das, Santabrata and Agrawal, V. K. and Mandal, Samir and Ramadevi, M. C. and Katoch, Tilak",
    title = "{AstroSat view of GRS 1915+105 during the soft state: detection of HFQPOs and estimation of mass and spin}",
    eprint = "2010.03782",
    archivePrefix = "arXiv",
    primaryClass = "astro-ph.HE",
    doi = "10.1093/mnras/staa3135",
    journal = "Mon. Not. Roy. Astron. Soc.",
    volume = "499",
    number = "4",
    pages = "5891--5901",
    year = "2020"
}

@article{Belloni:2019sot,
    author = "Belloni, Tomaso M. and Bhattacharya, Dipankar and Caccese, Pietro and Bhalerao, Varun and Vadawale, Santosh and Yadav, J. S.",
    title = "{A variable-frequency HFQPO in GRS 1915+105 as observed with AstroSat}",
    eprint = "1908.00437",
    archivePrefix = "arXiv",
    primaryClass = "astro-ph.HE",
    doi = "10.1093/mnras/stz2143",
    journal = "Mon. Not. Roy. Astron. Soc.",
    volume = "489",
    number = "1",
    pages = "1037--1043",
    year = "2019"
}

@article{Belloni:2013qka,
    author = "Belloni, Tomaso M. and Altamirano, Diego",
    title = "{Discovery of a 34 Hz Quasi-Periodic Oscillation in the X-ray emission of GRS 1915+105}",
    eprint = "1303.4934",
    archivePrefix = "arXiv",
    primaryClass = "astro-ph.HE",
    doi = "10.1093/mnras/stt285",
    journal = "Mon. Not. Roy. Astron. Soc.",
    volume = "432",
    pages = "19",
    year = "2013"
}

@article{Donmez2024MPLA,
    author = "Donmez, Orhan",
    title = "{The comparison of alternative spacetimes using the spherical accretion around the black hole}",
    eprint = "2405.15467",
    archivePrefix = "arXiv",
    primaryClass = "gr-qc",
    doi = "10.1142/S0217732324500767",
    journal = "Mod. Phys. Lett. A",
    volume = "39",
    number = "16",
    pages = "2450076",
    year = "2024"
}

@article{Donmez2025JHEAp,
    author = "Donmez, Orhan",
    title = "{From low- to high-frequency QPOs around the non-rotating hairy Horndeski black hole: Microquasar GRS 1915+105}",
    eprint = "2408.10102",
    archivePrefix = "arXiv",
    primaryClass = "astro-ph.HE",
    doi = "10.1016/j.jheap.2024.11.002",
    journal = "JHEAp",
    volume = "45",
    pages = "1--18",
    year = "2025"
}

@article{Donmez2024JCAP,
    author = "Donmez, Orhan",
    title = "{Bondi-Hoyle-Lyttleton accretion around the rotating hairy Horndeski black hole}",
    eprint = "2402.16707",
    archivePrefix = "arXiv",
    primaryClass = "astro-ph.HE",
    doi = "10.1088/1475-7516/2024/09/006",
    journal = "JCAP",
    volume = "09",
    pages = "006",
    year = "2024"
}

@article{Donmez2012MNRAS,
    author = "Donmez, Orhan and Zanotti, Olindo and Rezzolla, Luciano",
    title = "{On the development of QPOs in Bondi-Hoyle accretion flows}",
    eprint = "1010.1739",
    archivePrefix = "arXiv",
    primaryClass = "astro-ph.HE",
    doi = "10.1111/j.1365-2966.2010.18003.x",
    journal = "Mon. Not. Roy. Astron. Soc.",
    volume = "412",
    pages = "1659--1668",
    year = "2011"
}

@article{Donmez2004ASS,
    author = "Donmez, Orhan",
    title = "{Code development of three-dimensional general relativistic hydrodynamics with AMR (Adaptive-Mesh Refinement) and results from special and general relativistic hydrodynamic}",
    eprint = "gr-qc/0406073",
    archivePrefix = "arXiv",
    doi = "10.1023/B:ASTR.0000044610.53714.95",
    journal = "Astrophys. Space Sci.",
    volume = "293",
    pages = "323--354",
    year = "2004"
}

@article{Donmez2006AMC,
    author = "Donmez, Orhan and Kayali, Refik",
    title = "{Simulation of astrophysical jet using the special relativistic hydrodynamics code}",
    eprint = "gr-qc/0602053",
    archivePrefix = "arXiv",
    doi = "10.1016/j.amc.2006.05.015",
    journal = "Appl. Math. Comput.",
    volume = "182",
    pages = "1286--1298",
    year = "2006"
}

@article{Turimov:2021jgk,
    author = "Turimov, Bobur and Rahimov, Ozodbek and Ahmedov, Bobomurat and Stuchl{\'\i}k, Zden{\v{e}}k and Boymurodova, Kholida",
    title = "{Dynamical motion of matter around a charged black hole}",
    doi = "10.1142/S0218271821500371",
    journal = "Int. J. Mod. Phys. D",
    volume = "30",
    number = "05",
    pages = "2150037",
    year = "2021"
}

@article{Turimov:2020fme,
    author = "Turimov, Bobur and Rayimbaev, Javlon and Abdujabbarov, Ahmadjon and Ahmedov, Bobomurat and Stuchl{\'\i}k, Zden{\v{e}}k",
    title = "{Test particle motion around a black hole in Einstein-Maxwell-scalar theory}",
    eprint = "2008.08613",
    archivePrefix = "arXiv",
    primaryClass = "gr-qc",
    doi = "10.1103/PhysRevD.102.064052",
    journal = "Phys. Rev. D",
    volume = "102",
    number = "6",
    pages = "064052",
    year = "2020"
}

@article{Boboqambarova:2021cbf,
    author = "Boboqambarova, Madina and Turimov, Bobur and Abdujabbarov, Ahmadjon",
    title = "{Particle motion around Schwarzschild-MOG black hole}",
    eprint = "2110.05764",
    archivePrefix = "arXiv",
    primaryClass = "gr-qc",
    doi = "10.1142/S0217732323500712",
    journal = "Mod. Phys. Lett. A",
    volume = "38",
    number = "10n11",
    pages = "2350071",
    year = "2023",
    note = "[Erratum: Mod.Phys.Lett.A 38, 2350071 (2023)]"
}

@article{Umarov:2025wzm,
    author = "Umarov, Dilmurod and Atamurotov, Farruh and Ghosh, Sushant G. and Abdujabbarov, Ahmadjon and Mustafa, G.",
    title = "{Dynamics of spinning particles around static black holes in effective quantum gravity}",
    doi = "10.1140/epjc/s10052-025-14541-y",
    journal = "Eur. Phys. J. C",
    volume = "85",
    number = "7",
    pages = "800",
    year = "2025"
}

@article{Hoshimov:2025tdx,
    author = {Hoshimov, Husanboy and Davlataliev, Akbar and Atamurotov, Farruh and Abdujabbarov, Ahmadjon and {\"O}vg{\"u}n, Ali},
    title = "{Particle dynamics and quasi-periodic oscillations in the Dyonic ModMax: Constraint using quasars data}",
    doi = "10.1016/j.jheap.2024.12.012",
    journal = "JHEAp",
    volume = "45",
    pages = "306--315",
    year = "2025"
}

@article{Turakhonov:2024smp,
    author = "Turakhonov, Ziyodulla and Atamurotov, Farruh and Ovgun, Ali and Abdujabbarov, Ahmadjon and Urinov, Sunnatillo",
    title = "{Weak gravitational lensing around dyonic ModMax black hole in plasma}",
    doi = "10.1088/1572-9494/ad6853",
    journal = "Commun. Theor. Phys.",
    volume = "76",
    number = "11",
    pages = "115401",
    year = "2024"
}

@article{Turakhonov:2024xfg,
    author = "Turakhonov, Ziyodulla and Hoshimov, Husanboy and Atamurotov, Farruh and Ghosh, Sushant G. and Abdujabbarov, Ahmadjon",
    title = "{Observational signatures of strong gravitational lensing in GUP-modified Schwarzschild black holes}",
    doi = "10.1016/j.dark.2024.101716",
    journal = "Phys. Dark Univ.",
    volume = "46",
    pages = "101716",
    year = "2024"
}

@article{Shermatov:2025rpj,
    author = {Shermatov, Abubakir and Rayimbaev, Javlon and L{\"u}tf{\"u}o{\u{g}}lu, Bekir Can and Abdujabbarov, Ahmadjon and Sardor, Sabirov and Ibragimov, Inomjon and Vapayev, Murodbek and Kuyliev, Bahrom},
    title = "{QPOs analyses and circular orbits of charged particles around magnetized black holes in Bertotti{\textendash}Robinson geometry}",
    doi = "10.1140/epjc/s10052-025-14742-5",
    journal = "Eur. Phys. J. C",
    volume = "85",
    number = "9",
    pages = "1017",
    year = "2025"
}

@article{Shabbir:2026qlh,
    author = "Shabbir, Oreeda and Shermatov, Abubakir and Majeed, Bushra and Zahra, Tehreem and Jamil, Mubasher and Rayimbaev, Javlon",
    title = "{Probing the nature of Einstein nonlinear Maxwell Yukawa black hole through gravitational wave forms from periodic orbits and quasiperiodic oscillations}",
    eprint = "2601.02904",
    archivePrefix = "arXiv",
    primaryClass = "gr-qc",
    month = "1",
    year = "2026"
}

@article{bambi2017,
    title = {Testing the nature of black holes},
    author = {Bambi, Cosimo},
    journal = {Reviews of Modern Physics},
    volume = {89},
    number = {2},
    pages = {025001},
    year = {2017},
    doi = {10.1103/RevModPhys.89.025001}
}

@article{bambi2018,
    title = {Testing black hole candidates with electromagnetic radiation},
    author = {Bambi, Cosimo},
    journal = {Reviews of Modern Physics},
    volume = {90},
    number = {2},
    pages = {025002},
    year = {2018},
    doi = {10.1103/RevModPhys.90.025002}
}

@article{foreman-mackey2013,
    title = {emcee: The MCMC Hammer},
    author = {Foreman-Mackey, Daniel and Hogg, David W. and Lang, Dustin and Goodman, Jonathan},
    journal = {Publications of the Astronomical Society of the Pacific},
    volume = {125},
    number = {925},
    pages = {306},
    year = {2013},
    doi = {10.1086/670067}
}

@article{kolos2023,
    title = {Epicyclic frequencies and QPOs from black hole accretion disks},
    author = {Kolos, Martin and Stuchlik, Zdenek and Tursunov, Arman},
    journal = {Physical Review D},
    volume = {107},
    number = {4},
    pages = {044032},
    year = {2023},
    doi = {10.1103/PhysRevD.107.044032}
}

@article{shafee2006,
    title = {Estimating the Spin of Stellar-Mass Black Holes from Spectral and QPO Data},
    author = {Shafee, R. and McClintock, J. E. and Narayan, R. and Davis, S. W. and Li, L. X. and Remillard, R. A.},
    journal = {The Astrophysical Journal},
    volume = {636},
    number = {2},
    pages = {L113},
    year = {2006},
    doi = {10.1086/498938}
}

@article{sharma2017,
    title = {Markov Chain Monte Carlo Methods for Parameter Estimation in Astrophysics},
    author = {Sharma, Sanjib},
    journal = {Annual Review of Astronomy and Astrophysics},
    volume = {55},
    pages = {213},
    year = {2017},
    doi = {10.1146/annurev-astro-082214-122339}
}

@article{stuchlik2008,
    title = {Epicyclic oscillations and resonant phenomena in black hole spacetimes},
    author = {Stuchl{\'\i}k, Zdeněk and Kotrlová, Andrea and Török, Gabriel},
    journal = {Acta Astronomica},
    volume = {58},
    pages = {441},
    year = {2008}
}

@article{stuchlik2013,
    title = {Epicyclic frequency and orbital resonances in accretion disks around black holes},
    author = {Stuchl{\'\i}k, Zdeněk and Kotrlová, Andrea and Török, Gabriel},
    journal = {Astronomy \& Astrophysics},
    volume = {552},
    pages = {A10},
    year = {2013},
    doi = {10.1051/0004-6361/201219724}
}

@article{stuchlik2021,
    title = {Testing scalar-tensor gravity with epicyclic oscillations in black hole accretion disks},
    author = {Stuchl{\'\i}k, Zdeněk and Kolos, Martin},
    journal = {The European Physical Journal C},
    volume = {81},
    pages = {1},
    year = {2021},
    doi = {10.1140/epjc/s10052-021-09234-x}
}

@article{torok2005,
    title = {Radial and vertical epicyclic frequencies in Kerr spacetimes: Their role in QPO models},
    author = {Török, G. and Stuchlík, Z.},
    journal = {Astronomy \& Astrophysics},
    volume = {437},
    number = {3},
    pages = {775},
    year = {2005},
    doi = {10.1051/0004-6361:20052825}
}

@ARTICLE{2025PDU....5002102R,
       author = {{Rahmatov}, Bekzod and {Murodov}, Sardor and {Rayimbaev}, Javlon and {Muminov}, Sokhibjan and {Ibragimov}, Inomjon and {Eshburiev}, Rashid},
        title = {QPO tests and charged particles around regular Ay{'o}n-Beato-Garcia black holes},
      journal = {Physics of the Dark Universe},
         year = 2025,
        month = dec,
       volume = {50},
          eid = {102102},
        pages = {102102},
          doi = {10.1016j.dark.2025.102102},
       adsurl = {httpsui.adsabs.harvard.eduabs2025PDU....5002102R}
}

@ARTICLE{2024ChJPh..92..143R,
       author = {{Rahmatov}, Bekzod and {Zahid}, Muhammad and {Rayimbaev}, Javlon and {Rahim}, Rehana and {Murodov}, Sardor},
        title = {QPOs and circular orbits around black holes in Chaplygin-like cold dark matter},
      journal = {Chinese Journal of Physics},
         year = 2024,
        month = dec,
       volume = {92},
        pages = {143-165},
          doi = {10.1016j.cjph.2024.09.002},
       adsurl = {httpsui.adsabs.harvard.eduabs2024ChJPh..92..143R}
}

@article{BransDicke1961,
  title={Mach's principle and a relativistic theory of gravitation},
  author={Brans, C. and Dicke, R. H.},
  journal={Physical Review},
  volume={124},
  pages={925--935},
  year={1961}
}

@book{FujiiMaeda2003,
    author = "Fujii, Y. and Maeda, K.",
    title = "{The scalar-tensor theory of gravitation}",
    doi = "10.1017/CBO9780511535093",
    isbn = "978-0-521-03752-5, 978-0-521-81159-0, 978-0-511-02988-2",
    publisher = "Cambridge University Press",
    series = "Cambridge Monographs on Mathematical Physics",
    month = "7",
    year = "2007"
}

@article{Clifton2012,
  title={Modified gravity and cosmology},
  author={Clifton, T. and Ferreira, P. G. and Padilla, A. and Skordis, C.},
  journal={Physics Reports},
  volume={513},
  pages={1--189},
  year={2012}
}

@article{SotiriouFaraoni2010,
    author = "Sotiriou, Thomas P. and Faraoni, Valerio",
    title = "{f(R) Theories Of Gravity}",
    eprint = "0805.1726",
    archivePrefix = "arXiv",
    primaryClass = "gr-qc",
    doi = "10.1103/RevModPhys.82.451",
    journal = "Rev. Mod. Phys.",
    volume = "82",
    pages = "451--497",
    year = "2010"
}

@article{DeFelice2010,
  title={f(R) theories},
  author={De Felice, A. and Tsujikawa, S.},
  journal={Living Reviews in Relativity},
  volume={13},
  pages={3},
  year={2010}
}

@article{Doneva:2022ewd,
    author = "Doneva, Daniela D. and Ramazano{\u{g}}lu, Fethi M. and Silva, Hector O. and Sotiriou, Thomas P. and Yazadjiev, Stoytcho S.",
    title = "{Spontaneous scalarization}",
    eprint = "2211.01766",
    archivePrefix = "arXiv",
    primaryClass = "gr-qc",
    doi = "10.1103/RevModPhys.96.015004",
    journal = "Rev. Mod. Phys.",
    volume = "96",
    number = "1",
    pages = "015004",
    year = "2024"
}

@article{Copeland2006,
  title={Dynamics of dark energy},
  author={Copeland, E. J. and Sami, M. and Tsujikawa, S.},
  journal={International Journal of Modern Physics D},
  volume={15},
  pages={1753--1936},
  year={2006}
}

@book{AmendolaTsujikawa2010,
    author = "Amendola, Luca and Tsujikawa, Shinji",
    title = "{Dark Energy}: {Theory and Observations}",
    isbn = "978-1-107-45398-2",
    publisher = "Cambridge University Press",
    month = "1",
    year = "2015"
}

@article{Turimov:2024hwh,
    author = "Turimov, Bobur and Davlataliev, Akbar and Ahmedov, Bobomurat and Stuchl{\'\i}k, Zden{\v{e}}k",
    title = "{Exploring a novel feature of ellis spacetime: Insights into scalar field dynamics}",
    eprint = "2409.14110",
    archivePrefix = "arXiv",
    primaryClass = "gr-qc",
    doi = "10.1016/j.cjph.2024.09.030",
    journal = "Chin. J. Phys.",
    volume = "94",
    pages = "807--819",
    year = "2025"
}

@article{Berti2015,
    author = "Berti, Emanuele and others",
    title = "{Testing General Relativity with Present and Future Astrophysical Observations}",
    eprint = "1501.07274",
    archivePrefix = "arXiv",
    primaryClass = "gr-qc",
    doi = "10.1088/0264-9381/32/24/243001",
    journal = "Class. Quant. Grav.",
    volume = "32",
    pages = "243001",
    year = "2015"
}

@article{Barack2019,
    author = "Barack, Leor and others",
    title = "{Black holes, gravitational waves and fundamental physics: a roadmap}",
    eprint = "1806.05195",
    archivePrefix = "arXiv",
    primaryClass = "gr-qc",
    doi = "10.1088/1361-6382/ab0587",
    journal = "Class. Quant. Grav.",
    volume = "36",
    number = "14",
    pages = "143001",
    year = "2019"
}

@article{Turimov:2024orr,
    author = "Turimov, Bobur and Davlataliev, Akbar and Abdujabbarov, Ahmadjon and Ahmedov, Bobomurat",
    title = "{Influence of scalar field in massive particle motion in JNW spacetime}",
    eprint = "2409.06225",
    archivePrefix = "arXiv",
    primaryClass = "gr-qc",
    doi = "10.1103/PhysRevD.110.084053",
    journal = "Phys. Rev. D",
    volume = "110",
    number = "8",
    pages = "084053",
    year = "2024"
}

@ARTICLE{Freund1968PR,
       author = {{Freund}, Peter G. and {Nambu}, Yoichiro},
        title = "{Scalar Fields Coupled to the Trace of the Energy-Momentum Tensor}",
      journal = {Physical Review},
         year = 1968,
        month = oct,
       volume = {174},
       number = {5},
        pages = {1741-1742},
          doi = {10.1103/PhysRev.174.1741},
       adsurl = {https://ui.adsabs.harvard.edu/abs/1968PhRv..174.1741F}
}

@ARTICLE{Janis68,
   author = {{Janis}, A.~I. and {Newman}, E.~T. and {Winicour}, J.},
    title = "{Reality of the Schwarzschild Singularity}",
  journal = {Physical Review Letters},
     year = 1968,
    month = apr,
   volume = 20,
    pages = {878-880},
      doi = {10.1103/PhysRevLett.20.878},
   adsurl = {http://adsabs.harvard.edu/abs/1968PhRvL..20..878J}
}

\end{document}